\AtBeginDocument{\emergencystretch=1em}

  \documentclass[sigplan,10pt,nonacm]{acmart}
\setcopyright{none}

\usepackage{booktabs}
\usepackage{graphicx}
\usepackage{subcaption}
\usepackage{amsmath}
\usepackage{fancyvrb}
\usepackage{tikz}
\usepackage{xcolor}
\usetikzlibrary{arrows.meta,positioning,calc,fit,backgrounds,shapes.geometric,decorations.markings,patterns.meta}
\definecolor{cfBlue}{HTML}{2F6FB5}
\definecolor{cfRed}{HTML}{C0392B}
\definecolor{cfGreen}{HTML}{2E7D50}
\definecolor{cfAmber}{HTML}{B4711A}
\definecolor{cfPurple}{HTML}{6A4C93}
\definecolor{cfAccent}{HTML}{A61B1B}
\colorlet{icoS}{black!72}   % icon stroke
\colorlet{icoF}{black!10}   % icon fill
\tikzset{
  ilw/.style={line width=0.35pt},
  pics/ico server/.style={code={
    \foreach \y in {-0.17,0,0.17}{%
      \draw[ilw, rounded corners=0.6pt, fill=icoF, draw=icoS]
        (-0.21,\y-0.068) rectangle (0.21,\y+0.068);
      \fill[icoS] (-0.15,\y) circle (0.021);}
  }},
  pics/ico queue/.style={code={
    \draw[ilw, draw=icoS]
      (-0.27,0.18) -- (-0.27,-0.17) -- (0.27,-0.17) -- (0.27,0.18);
    \foreach \x in {-0.15,0,0.15}
      \draw[ilw, rounded corners=0.5pt, fill=black!18, draw=icoS]
        (\x-0.055,-0.10) rectangle (\x+0.055,0.05);
  }},
  pics/ico gear/.style={code={
    \foreach \a in {0,45,...,315}
      \fill[icoS, rotate=\a] (-0.035,0.11) rectangle (0.035,0.215);
    \draw[ilw, fill=black!18, draw=icoS] (0,0) circle (0.155);
    \fill[white] (0,0) circle (0.058);
  }},
  pics/ico flask/.style={code={
    \draw[ilw, fill=white, draw=icoS]
      (-0.075,0.21) -- (-0.075,0.06) -- (-0.23,-0.19)
      .. controls (-0.25,-0.23) and (-0.20,-0.25) .. (-0.15,-0.25)
      -- (0.15,-0.25)
      .. controls (0.20,-0.25) and (0.25,-0.23) .. (0.23,-0.19)
      -- (0.075,0.06) -- (0.075,0.21);
    \draw[ilw, draw=icoS] (-0.11,0.21) -- (0.11,0.21);
    \path[fill=black!18]
      (-0.17,-0.12) -- (0.17,-0.12) -- (0.23,-0.21)
      .. controls (0.24,-0.23) and (0.20,-0.25) .. (0.15,-0.25)
      -- (-0.15,-0.25)
      .. controls (-0.20,-0.25) and (-0.24,-0.23) .. (-0.23,-0.21) -- cycle;
    \draw[ilw, draw=icoS] (-0.17,-0.12) -- (0.17,-0.12);
  }},
  pics/ico depot/.style={code={   % warehouse: low roof, three bays
    \draw[ilw, fill=icoF,  draw=icoS] (-0.27,-0.20) rectangle (0.27,0.07);
    \draw[ilw, fill=black!20, draw=icoS]
      (-0.30,0.07) -- (-0.19,0.21) -- (0.19,0.21) -- (0.30,0.07) -- cycle;
    \foreach \x in {-0.155,0,0.155}
      \draw[ilw, fill=white, draw=icoS] (\x-0.055,-0.20) rectangle (\x+0.055,-0.05);
  }},
  pics/ico truck/.style={code={   % delivery van, side view, facing right
    \draw[ilw, fill=icoF, draw=icoS, rounded corners=0.8pt] (-0.30,-0.05) rectangle (0.05,0.21);
    \draw[ilw, fill=icoF, draw=icoS]
      (0.05,-0.05) -- (0.05,0.21) -- (0.16,0.21) -- (0.30,0.07) -- (0.30,-0.05) -- cycle;
    \draw[ilw, fill=white, draw=icoS]
      (0.09,0.17) -- (0.15,0.17) -- (0.255,0.065) -- (0.09,0.065) -- cycle;
    \draw[ilw, draw=icoS] (-0.30,0.06) -- (0.05,0.06);
    \draw[ilw, fill=white, draw=icoS] (-0.17,-0.10) circle (0.083);
    \draw[ilw, fill=white, draw=icoS] ( 0.18,-0.10) circle (0.083);
    \fill[icoS] (-0.17,-0.10) circle (0.031);
    \fill[icoS] ( 0.18,-0.10) circle (0.031);
  }},
  pics/ico home/.style={code={
    \draw[ilw, fill=icoF,  draw=icoS] (-0.19,-0.20) rectangle (0.19,0.04);
    \draw[ilw, fill=black!20, draw=icoS] (-0.25,0.04) -- (0,0.22) -- (0.25,0.04) -- cycle;
    \draw[ilw, fill=white,    draw=icoS] (-0.06,-0.20) rectangle (0.06,-0.03);
    \fill[black!45] (0.11,0.10) rectangle (0.16,0.22);
  }},
  pics/ico parcel/.style={code={   % shipping box: body, lid, tape
    \draw[ilw, fill=white, draw=icoS] (-0.195,-0.20) rectangle (0.195,0.075);
    \draw[ilw, fill=black!12, draw=icoS] (-0.225,0.075) rectangle (0.225,0.205);
    \draw[ilw, draw=icoS] (-0.05,0.075) -- (-0.05,0.205);
    \draw[ilw, draw=icoS] ( 0.05,0.075) -- ( 0.05,0.205);
    \draw[ilw, draw=icoS] (0,0.075) -- (0,-0.20);
  }},
  pics/ico req/.style={code={     % request in flight
    \draw[ilw, fill=white, draw=icoS]
      (-0.20,-0.145) -- (0.05,-0.145) -- (0.21,0) -- (0.05,0.145) -- (-0.20,0.145) -- cycle;
    \draw[ilw, draw=black!45] (-0.14,0.055) -- (0.01,0.055);
    \draw[ilw, draw=black!45] (-0.14,-0.055) -- (0.01,-0.055);
  }},
  pics/ico break/.style={code={   % snapped chain link
    \draw[line width=0.7pt, cfRed] (-0.155,0) ellipse (0.075 and 0.135);
    \draw[line width=0.7pt, cfRed] ( 0.155,0) ellipse (0.075 and 0.135);
    \foreach \a in {55,90,125}
      \draw[line width=0.45pt, cfRed] (\a:0.09) -- (\a:0.20);
  }},
  pics/ico root/.style={code={    % "new root": flag on a stem
    \draw[line width=0.5pt, cfRed] (-0.10,-0.16) -- (-0.10,0.18);
    \draw[ilw, fill=cfRed!25, draw=cfRed] (-0.10,0.18) -- (0.16,0.10) -- (-0.10,0.02) -- cycle;
  }},
  pics/ico pulse/.style={code={   % controlled traffic generator
    \draw[ilw, rounded corners=1.2pt, fill=white, draw=icoS]
      (-0.30,-0.20) rectangle (0.30,0.20);
    \draw[line width=0.6pt, black!75]
      (-0.24,-0.02) -- (-0.14,-0.02) -- (-0.10,0.12) -- (-0.04,-0.13)
      -- (0.02,0.12) -- (0.08,-0.13) -- (0.13,-0.02) -- (0.24,-0.02);
  }},
  pics/ico flat/.style={code={    % idle: no request traffic
    \draw[ilw, rounded corners=1.2pt, fill=white, draw=icoS]
      (-0.30,-0.20) rectangle (0.30,0.20);
    \draw[line width=0.6pt, black!45] (-0.24,-0.02) -- (0.24,-0.02);
  }},
  pics/ico wave/.style={code={    % bounded repair wave over services
    \foreach \x in {-0.20,0,0.20}
      \draw[ilw, rounded corners=0.5pt, fill=icoF, draw=icoS]
        (\x-0.07,-0.21) rectangle (\x+0.07,0.02);
    \draw[line width=0.7pt, black!75, -{Stealth[length=1.1mm]}]
      (-0.24,0.11) .. controls (-0.08,0.24) and (0.08,-0.02) .. (0.25,0.13);
  }},
  pics/ico fleet/.style={code={   % the fleet / visible repair set
    \foreach \x in {-0.18,0,0.18}
      \foreach \y in {-0.18,0,0.18}
        \draw[ilw, rounded corners=0.4pt, fill=icoF, draw=icoS]
          (\x-0.062,\y-0.062) rectangle (\x+0.062,\y+0.062);
    \foreach \p in {(-0.18,0.18),(0.18,0),(0,-0.18)}
      \draw[line width=0.8pt, rounded corners=0.4pt, fill=cfAccent!25, draw=cfAccent] \p ++(-0.062,-0.062)
        rectangle ++(0.124,0.124);
  }},
  pics/ico bot/.style={code={      % repair agent
    \draw[ilw, draw=icoS] (0,0.18) -- (0,0.27);
    \draw[ilw, fill=black!18, draw=icoS] (0,0.29) circle (0.025);
    \draw[ilw, rounded corners=1.2pt, fill=icoF, draw=icoS]
      (-0.22,-0.15) rectangle (0.22,0.18);
    \draw[ilw, fill=black!18, draw=icoS]
      (-0.27,-0.07) rectangle (-0.22,0.08);
    \draw[ilw, fill=black!18, draw=icoS]
      (0.22,-0.07) rectangle (0.27,0.08);
    \fill[icoS] (-0.09,0.05) circle (0.030);
    \fill[icoS] (0.09,0.05) circle (0.030);
    \draw[line width=0.45pt, draw=icoS, line cap=round]
      (-0.09,-0.065) -- (0.09,-0.065);
  }},
  pics/ico glass/.style={code={%  Font Awesome Free 6 (CC BY 4.0)
    \fill[icoS]
    (0.1415,0.0443) .. controls (0.1415,0.0033) and (0.1282,-0.0346) .. (0.1058,-0.0653)
      -- (0.2188,-0.1784)
      .. controls (0.2300,-0.1896) and (0.2300,-0.2077) .. (0.2188,-0.2188)
      .. controls (0.2076,-0.2300) and (0.1895,-0.2300) .. (0.1783,-0.2188)
      -- (0.0653,-0.1057)
      .. controls (0.0346,-0.1282) and (-0.0033,-0.1414) .. (-0.0442,-0.1414)
      .. controls (-0.1468,-0.1414) and (-0.2300,-0.0583) .. (-0.2300,0.0443)
      .. controls (-0.2300,0.1469) and (-0.1468,0.2300) .. (-0.0442,0.2300)
      .. controls (0.0583,0.2300) and (0.1415,0.1469) .. (0.1415,0.0443) -- cycle
    (-0.0442,-0.0843) .. controls (0.0268,-0.0843) and (0.0843,-0.0267) .. (0.0843,0.0443)
      .. controls (0.0843,0.1153) and (0.0268,0.1729) .. (-0.0442,0.1729)
      .. controls (-0.1152,0.1729) and (-0.1728,0.1153) .. (-0.1728,0.0443)
      .. controls (-0.1728,-0.0267) and (-0.1152,-0.0843) .. (-0.0442,-0.0843) -- cycle;
  }},
  pics/ico target/.style={code={%  Font Awesome Free 6 (CC BY 4.0)
    \fill[icoS]
    (0.0000,0.2300) .. controls (0.0159,0.2300) and (0.0288,0.2172) .. (0.0288,0.2013)
      -- (0.0288,0.1919) .. controls (0.1129,0.1794) and (0.1794,0.1128) .. (0.1919,0.0288)
      -- (0.2013,0.0288) .. controls (0.2172,0.0288) and (0.2300,0.0159) .. (0.2300,-0.0000)
      .. controls (0.2300,-0.0159) and (0.2172,-0.0288) .. (0.2013,-0.0288)
      -- (0.1919,-0.0288)
      .. controls (0.1794,-0.1129) and (0.1128,-0.1794) .. (0.0288,-0.1919)
      -- (0.0288,-0.2013)
      .. controls (0.0288,-0.2172) and (0.0159,-0.2300) .. (0.0000,-0.2300)
      .. controls (-0.0159,-0.2300) and (-0.0288,-0.2172) .. (-0.0288,-0.2013)
      -- (-0.0288,-0.1919)
      .. controls (-0.1129,-0.1794) and (-0.1794,-0.1129) .. (-0.1919,-0.0288)
      -- (-0.2013,-0.0288)
      .. controls (-0.2172,-0.0288) and (-0.2300,-0.0159) .. (-0.2300,-0.0000)
      .. controls (-0.2300,0.0159) and (-0.2172,0.0288) .. (-0.2013,0.0288)
      -- (-0.1919,0.0288)
      .. controls (-0.1794,0.1129) and (-0.1129,0.1794) .. (-0.0288,0.1919)
      -- (-0.0288,0.2013)
      .. controls (-0.0288,0.2172) and (-0.0159,0.2300) .. (0.0000,0.2300) -- cycle
    (-0.1335,-0.0288)
      .. controls (-0.1223,-0.0811) and (-0.0810,-0.1223) .. (-0.0288,-0.1335)
      -- (-0.0288,-0.1150)
      .. controls (-0.0288,-0.0991) and (-0.0159,-0.0863) .. (0.0000,-0.0863)
      .. controls (0.0159,-0.0863) and (0.0288,-0.0991) .. (0.0288,-0.1150)
      -- (0.0288,-0.1335)
      .. controls (0.0811,-0.1223) and (0.1223,-0.0810) .. (0.1335,-0.0288)
      -- (0.1150,-0.0288)
      .. controls (0.0991,-0.0288) and (0.0863,-0.0159) .. (0.0863,-0.0000)
      .. controls (0.0863,0.0159) and (0.0991,0.0288) .. (0.1150,0.0288) -- (0.1335,0.0288)
      .. controls (0.1223,0.0811) and (0.0811,0.1223) .. (0.0288,0.1335) -- (0.0288,0.1150)
      .. controls (0.0288,0.0991) and (0.0159,0.0863) .. (0.0000,0.0863)
      .. controls (-0.0159,0.0863) and (-0.0288,0.0991) .. (-0.0288,0.1150)
      -- (-0.0288,0.1335)
      .. controls (-0.0811,0.1223) and (-0.1223,0.0811) .. (-0.1335,0.0288)
      -- (-0.1150,0.0288)
      .. controls (-0.0991,0.0288) and (-0.0863,0.0159) .. (-0.0863,-0.0000)
      .. controls (-0.0863,-0.0159) and (-0.0991,-0.0288) .. (-0.1150,-0.0288)
      -- (-0.1335,-0.0288) -- cycle
    (0.0000,0.0288) .. controls (0.0159,0.0288) and (0.0288,0.0159) .. (0.0288,-0.0000)
      .. controls (0.0288,-0.0159) and (0.0159,-0.0288) .. (0.0000,-0.0288)
      .. controls (-0.0159,-0.0288) and (-0.0288,-0.0159) .. (-0.0288,-0.0000)
      .. controls (-0.0288,0.0159) and (-0.0159,0.0288) .. (0.0000,0.0288) -- cycle;
  }},
  pics/ico wrench/.style={code={%  Font Awesome Free 6 (CC BY 4.0)
    \fill[icoS]
    (0.0863,-0.0575) .. controls (0.1657,-0.0575) and (0.2300,0.0068) .. (0.2300,0.0863)
      .. controls (0.2300,0.1000) and (0.2280,0.1133) .. (0.2244,0.1260)
      .. controls (0.2216,0.1357) and (0.2097,0.1378) .. (0.2026,0.1307) -- (0.1336,0.0617)
      .. controls (0.1309,0.0590) and (0.1272,0.0575) .. (0.1234,0.0575) -- (0.0719,0.0575)
      .. controls (0.0640,0.0575) and (0.0575,0.0640) .. (0.0575,0.0719) -- (0.0575,0.1234)
      .. controls (0.0575,0.1272) and (0.0590,0.1309) .. (0.0617,0.1336) -- (0.1307,0.2026)
      .. controls (0.1378,0.2097) and (0.1356,0.2216) .. (0.1260,0.2244)
      .. controls (0.1133,0.2280) and (0.1000,0.2300) .. (0.0863,0.2300)
      .. controls (0.0068,0.2300) and (-0.0575,0.1657) .. (-0.0575,0.0863)
      .. controls (-0.0575,0.0691) and (-0.0544,0.0526) .. (-0.0490,0.0373)
      -- (-0.2121,-0.1259)
      .. controls (-0.2235,-0.1373) and (-0.2300,-0.1528) .. (-0.2300,-0.1690)
      .. controls (-0.2300,-0.2027) and (-0.2027,-0.2300) .. (-0.1690,-0.2300)
      .. controls (-0.1528,-0.2300) and (-0.1373,-0.2235) .. (-0.1259,-0.2121)
      -- (0.0373,-0.0490)
      .. controls (0.0526,-0.0545) and (0.0691,-0.0575) .. (0.0863,-0.0575) -- cycle
    (-0.1581,-0.1366)
      .. controls (-0.1462,-0.1366) and (-0.1366,-0.1462) .. (-0.1366,-0.1581)
      .. controls (-0.1366,-0.1700) and (-0.1462,-0.1797) .. (-0.1581,-0.1797)
      .. controls (-0.1700,-0.1797) and (-0.1797,-0.1700) .. (-0.1797,-0.1581)
      .. controls (-0.1797,-0.1462) and (-0.1700,-0.1366) .. (-0.1581,-0.1366) -- cycle;
  }},
  pics/ico scale/.style={code={%  Font Awesome Free 6 (CC BY 4.0)
    \fill[icoS]
    (0.0462,0.1604) -- (0.1379,0.1604)
      .. controls (0.1506,0.1604) and (0.1608,0.1502) .. (0.1608,0.1375)
      .. controls (0.1608,0.1248) and (0.1506,0.1146) .. (0.1379,0.1146) -- (0.0565,0.1146)
      .. controls (0.0528,0.0961) and (0.0401,0.0809) .. (0.0233,0.0736) -- (0.0233,-0.1375)
      -- (0.1379,-0.1375)
      .. controls (0.1506,-0.1375) and (0.1608,-0.1477) .. (0.1608,-0.1604)
      .. controls (0.1608,-0.1731) and (0.1506,-0.1833) .. (0.1379,-0.1833)
      -- (0.0004,-0.1833) -- (-0.1371,-0.1833)
      .. controls (-0.1498,-0.1833) and (-0.1600,-0.1731) .. (-0.1600,-0.1604)
      .. controls (-0.1600,-0.1477) and (-0.1498,-0.1375) .. (-0.1371,-0.1375)
      -- (-0.0225,-0.1375) -- (-0.0225,0.0736)
      .. controls (-0.0394,0.0809) and (-0.0520,0.0962) .. (-0.0558,0.1146)
      -- (-0.1371,0.1146)
      .. controls (-0.1498,0.1146) and (-0.1600,0.1248) .. (-0.1600,0.1375)
      .. controls (-0.1600,0.1502) and (-0.1498,0.1604) .. (-0.1371,0.1604)
      -- (-0.0454,0.1604)
      .. controls (-0.0350,0.1743) and (-0.0184,0.1833) .. (0.0004,0.1833)
      .. controls (0.0192,0.1833) and (0.0358,0.1743) .. (0.0462,0.1604) -- cycle
    (0.0860,-0.0458) -- (0.1898,-0.0458) -- (0.1379,0.0431) -- (0.0860,-0.0458) -- cycle
    (0.1379,-0.1146) .. controls (0.0929,-0.1146) and (0.0554,-0.0902) .. (0.0477,-0.0581)
      .. controls (0.0458,-0.0502) and (0.0484,-0.0421) .. (0.0525,-0.0351)
      -- (0.1206,0.0818) .. controls (0.1242,0.0879) and (0.1308,0.0917) .. (0.1379,0.0917)
      .. controls (0.1450,0.0917) and (0.1516,0.0879) .. (0.1552,0.0818) -- (0.2233,-0.0351)
      .. controls (0.2274,-0.0421) and (0.2300,-0.0502) .. (0.2281,-0.0581)
      .. controls (0.2204,-0.0902) and (0.1829,-0.1146) .. (0.1379,-0.1146) -- cycle
    (-0.1380,0.0431) -- (-0.1898,-0.0458) -- (-0.0860,-0.0458) -- (-0.1380,0.0431) -- cycle
    (-0.2281,-0.0581)
      .. controls (-0.2300,-0.0502) and (-0.2274,-0.0421) .. (-0.2233,-0.0351)
      -- (-0.1552,0.0818)
      .. controls (-0.1516,0.0879) and (-0.1450,0.0917) .. (-0.1379,0.0917)
      .. controls (-0.1308,0.0917) and (-0.1242,0.0879) .. (-0.1206,0.0818)
      -- (-0.0525,-0.0351)
      .. controls (-0.0484,-0.0421) and (-0.0458,-0.0502) .. (-0.0477,-0.0581)
      .. controls (-0.0555,-0.0902) and (-0.0929,-0.1146) .. (-0.1380,-0.1146)
      .. controls (-0.1830,-0.1146) and (-0.2204,-0.0902) .. (-0.2281,-0.0581) -- cycle;
  }},
  pics/ico chart/.style={code={%  Font Awesome Free 6 (CC BY 4.0)
    \fill[icoS]
    (-0.2013,0.2013) .. controls (-0.1853,0.2013) and (-0.1725,0.1884) .. (-0.1725,0.1725)
      -- (-0.1725,-0.1294)
      .. controls (-0.1725,-0.1373) and (-0.1660,-0.1438) .. (-0.1581,-0.1438)
      -- (0.2013,-0.1438)
      .. controls (0.2172,-0.1438) and (0.2300,-0.1566) .. (0.2300,-0.1725)
      .. controls (0.2300,-0.1884) and (0.2172,-0.2013) .. (0.2013,-0.2013)
      -- (-0.1581,-0.2013)
      .. controls (-0.1978,-0.2013) and (-0.2300,-0.1691) .. (-0.2300,-0.1294)
      -- (-0.2300,0.1725)
      .. controls (-0.2300,0.1884) and (-0.2172,0.2013) .. (-0.2013,0.2013) -- cycle
    (-0.0863,0.0288) .. controls (-0.0703,0.0288) and (-0.0575,0.0159) .. (-0.0575,-0.0000)
      -- (-0.0575,-0.0575)
      .. controls (-0.0575,-0.0734) and (-0.0703,-0.0863) .. (-0.0863,-0.0863)
      .. controls (-0.1022,-0.0863) and (-0.1150,-0.0734) .. (-0.1150,-0.0575)
      -- (-0.1150,-0.0000)
      .. controls (-0.1150,0.0159) and (-0.1022,0.0288) .. (-0.0863,0.0288) -- cycle
    (0.0288,0.0863) -- (0.0288,-0.0575)
      .. controls (0.0288,-0.0734) and (0.0159,-0.0863) .. (0.0000,-0.0863)
      .. controls (-0.0159,-0.0863) and (-0.0288,-0.0734) .. (-0.0288,-0.0575)
      -- (-0.0288,0.0863)
      .. controls (-0.0288,0.1022) and (-0.0159,0.1150) .. (0.0000,0.1150)
      .. controls (0.0159,0.1150) and (0.0288,0.1022) .. (0.0288,0.0863) -- cycle
    (0.0863,0.0575) .. controls (0.1022,0.0575) and (0.1150,0.0447) .. (0.1150,0.0288)
      -- (0.1150,-0.0575)
      .. controls (0.1150,-0.0734) and (0.1022,-0.0863) .. (0.0863,-0.0863)
      .. controls (0.0703,-0.0863) and (0.0575,-0.0734) .. (0.0575,-0.0575)
      -- (0.0575,0.0288) .. controls (0.0575,0.0447) and (0.0703,0.0575) .. (0.0863,0.0575)
      -- cycle
    (0.2013,0.1438) -- (0.2013,-0.0575)
      .. controls (0.2013,-0.0734) and (0.1884,-0.0863) .. (0.1725,-0.0863)
      .. controls (0.1566,-0.0863) and (0.1438,-0.0734) .. (0.1438,-0.0575)
      -- (0.1438,0.1438) .. controls (0.1438,0.1597) and (0.1566,0.1725) .. (0.1725,0.1725)
      .. controls (0.1884,0.1725) and (0.2013,0.1597) .. (0.2013,0.1438) -- cycle;
  }},
  pics/ico funnel/.style={code={%  Font Awesome Free 6 (CC BY 4.0)
    \fill[icoS]
    (-0.2240,0.1796) .. controls (-0.2182,0.1921) and (-0.2057,0.2000) .. (-0.1920,0.2000)
      -- (0.1920,0.2000) .. controls (0.2058,0.2000) and (0.2183,0.1921) .. (0.2241,0.1796)
      .. controls (0.2300,0.1672) and (0.2282,0.1525) .. (0.2195,0.1419) -- (0.0569,-0.0568)
      -- (0.0569,-0.1698)
      .. controls (0.0569,-0.1805) and (0.0509,-0.1904) .. (0.0412,-0.1952)
      .. controls (0.0315,-0.2000) and (0.0200,-0.1990) .. (0.0114,-0.1925)
      -- (-0.0455,-0.1499)
      .. controls (-0.0527,-0.1445) and (-0.0568,-0.1361) .. (-0.0568,-0.1271)
      -- (-0.0568,-0.0568) -- (-0.2195,0.1420)
      .. controls (-0.2281,0.1525) and (-0.2300,0.1673) .. (-0.2240,0.1796) -- cycle;
  }},
  pics/ico eye/.style={code={%  Font Awesome Free 6 (CC BY 4.0)
    \fill[icoS]
    (-0.0000,0.1784) .. controls (-0.0644,0.1784) and (-0.1159,0.1491) .. (-0.1534,0.1142)
      .. controls (-0.1907,0.0796) and (-0.2156,0.0382) .. (-0.2274,0.0098)
      .. controls (-0.2300,0.0035) and (-0.2300,-0.0035) .. (-0.2274,-0.0098)
      .. controls (-0.2156,-0.0382) and (-0.1907,-0.0796) .. (-0.1534,-0.1142)
      .. controls (-0.1159,-0.1491) and (-0.0644,-0.1784) .. (-0.0000,-0.1784)
      .. controls (0.0643,-0.1784) and (0.1158,-0.1491) .. (0.1533,-0.1142)
      .. controls (0.1906,-0.0795) and (0.2155,-0.0382) .. (0.2274,-0.0098)
      .. controls (0.2300,-0.0035) and (0.2300,0.0035) .. (0.2274,0.0098)
      .. controls (0.2155,0.0382) and (0.1906,0.0796) .. (0.1533,0.1142)
      .. controls (0.1158,0.1491) and (0.0643,0.1784) .. (-0.0000,0.1784) -- cycle
    (-0.1147,-0.0000) .. controls (-0.1147,0.0633) and (-0.0634,0.1147) .. (-0.0000,0.1147)
      .. controls (0.0633,0.1147) and (0.1146,0.0633) .. (0.1146,0.0000)
      .. controls (0.1146,-0.0633) and (0.0633,-0.1147) .. (-0.0000,-0.1147)
      .. controls (-0.0634,-0.1147) and (-0.1147,-0.0633) .. (-0.1147,-0.0000) -- cycle
    (-0.0000,0.0510) .. controls (-0.0000,0.0229) and (-0.0229,-0.0000) .. (-0.0510,-0.0000)
      .. controls (-0.0567,-0.0000) and (-0.0621,0.0010) .. (-0.0672,0.0026)
      .. controls (-0.0715,0.0041) and (-0.0766,0.0014) .. (-0.0765,-0.0033)
      .. controls (-0.0762,-0.0088) and (-0.0754,-0.0143) .. (-0.0739,-0.0197)
      .. controls (-0.0630,-0.0605) and (-0.0211,-0.0847) .. (0.0197,-0.0738)
      .. controls (0.0605,-0.0629) and (0.0847,-0.0209) .. (0.0738,0.0198)
      .. controls (0.0649,0.0529) and (0.0357,0.0751) .. (0.0032,0.0764)
      .. controls (-0.0014,0.0766) and (-0.0041,0.0716) .. (-0.0027,0.0671)
      .. controls (-0.0010,0.0620) and (-0.0000,0.0566) .. (-0.0000,0.0510) -- cycle;
  }},
  pics/ico book/.style={code={%  Font Awesome Free 6 (CC BY 4.0)
    \fill[icoS]
    (-0.1150,0.2300) .. controls (-0.1626,0.2300) and (-0.2013,0.1914) .. (-0.2013,0.1438)
      -- (-0.2013,-0.1438)
      .. controls (-0.2013,-0.1914) and (-0.1626,-0.2300) .. (-0.1150,-0.2300)
      -- (0.1438,-0.2300) -- (0.1725,-0.2300)
      .. controls (0.1884,-0.2300) and (0.2013,-0.2172) .. (0.2013,-0.2013)
      .. controls (0.2013,-0.1853) and (0.1884,-0.1725) .. (0.1725,-0.1725)
      -- (0.1725,-0.1150)
      .. controls (0.1884,-0.1150) and (0.2013,-0.1022) .. (0.2013,-0.0863)
      -- (0.2013,0.2013) .. controls (0.2013,0.2172) and (0.1884,0.2300) .. (0.1725,0.2300)
      -- (0.1438,0.2300) -- (-0.1150,0.2300) -- cycle
    (-0.1150,-0.1150) -- (0.1150,-0.1150) -- (0.1150,-0.1725) -- (-0.1150,-0.1725)
      .. controls (-0.1309,-0.1725) and (-0.1438,-0.1597) .. (-0.1438,-0.1438)
      .. controls (-0.1438,-0.1278) and (-0.1309,-0.1150) .. (-0.1150,-0.1150) -- cycle
    (-0.0863,0.1006) .. controls (-0.0863,0.1085) and (-0.0798,0.1150) .. (-0.0719,0.1150)
      -- (0.1006,0.1150) .. controls (0.1085,0.1150) and (0.1150,0.1085) .. (0.1150,0.1006)
      .. controls (0.1150,0.0927) and (0.1085,0.0863) .. (0.1006,0.0863) -- (-0.0719,0.0863)
      .. controls (-0.0798,0.0863) and (-0.0863,0.0927) .. (-0.0863,0.1006) -- cycle
    (-0.0719,0.0575) -- (0.1006,0.0575)
      .. controls (0.1085,0.0575) and (0.1150,0.0510) .. (0.1150,0.0431)
      .. controls (0.1150,0.0352) and (0.1085,0.0288) .. (0.1006,0.0288) -- (-0.0719,0.0288)
      .. controls (-0.0798,0.0288) and (-0.0863,0.0352) .. (-0.0863,0.0431)
      .. controls (-0.0863,0.0510) and (-0.0798,0.0575) .. (-0.0719,0.0575) -- cycle;
  }},
  pics/ico check/.style={code={
    \draw[line width=0.9pt, icoS, line cap=round]
      (-0.16,0.02) -- (-0.04,-0.12) -- (0.19,0.16);
  }},
  pics/ico cross/.style={code={
    \draw[line width=1.3pt, cfAccent, line cap=round] (-0.105,-0.105) -- (0.105,0.105);
    \draw[line width=1.3pt, cfAccent, line cap=round] (-0.105,0.105) -- (0.105,-0.105);
  }},
}

\newcommand{\bcode}[2]{%
  \foreach \bx/\bw in {0.075/0.017,0.120/0.009,0.160/0.021,0.205/0.009,0.245/0.017}
    \fill[#2] ($(#1.west)+(\bx-\bw,-0.085)$) rectangle ($(#1.west)+(\bx+\bw,0.085)$);}

  \newcommand{\systemname}{Backstitch}
\newcommand{\papertitle}{\systemname: Restoring Request Causality Across a Production Microservice Fleet}

\author{Ziyue Dang}
\authornote{Equal contribution.}
\affiliation{%
  \institution{TikTok Inc.}
  \country{USA}}
\email{ziyue.dang@bytedance.com}

\author{Qiuyu Wu}
\authornotemark[1]
\affiliation{%
  \institution{TikTok Inc.}
  \country{USA}}
\email{wuqiuyu@bytedance.com}

\author{Haoyun Xu}
\authornote{Corresponding author.}
\affiliation{%
  \institution{TikTok Inc.}
  \country{USA}}
\email{haoyun.xu@bytedance.com}

\author{Tongjue Wang}
\affiliation{%
  \institution{TikTok Inc.}
  \country{USA}}
\email{tongjue.wang@bytedance.com}

\author{Yongqing Ling}
\affiliation{%
  \institution{TikTok Inc.}
  \country{USA}}
\email{lingyongqing.1@bytedance.com}

\author{Weihao Chen}
\affiliation{%
  \institution{Douyin Vision Co., Ltd.}
  \country{China}}
\email{chenweihao.123@bytedance.com}

\author{Guangming Luo}
\affiliation{%
  \institution{Douyin Vision Co., Ltd.}
  \country{China}}
\email{luoguangming.ivan@bytedance.com}

\newcommand{\sys}{\textsf{\systemname}}
\newcommand{\paragraphb}[1]{\noindent\textbf{#1}\quad}
\newcommand{\mainref}[1]{%
  \ref{#1}}

\begin{document}

  \title[\systemname]{\papertitle}
  \begin{abstract}
  A major video platform runs on thousands of microservices, each request
propagating a context so downstream work can be traced and governed. At
handoffs outside instrumented paths, e.g., custom queues and callbacks, the
payload continues but the context does not, and the request still succeeds
under existing tests. Such breaks are silent and widespread: 673 of 1,133
services carried at least one. \sys{}, a specialized agentic system, repairs them
using the surviving execution as its reference: replay determines whether a
suspicious call is request-correlated, source analysis reaches the responsible
handoff, a bounded change restores its contract, and the same replay validates
the fix. Repairs restore the causal chain without disturbing the work it
describes: breaks at 240 of the repaired calls fell from 90.46\% to 4.69\%, and
over 112 days the fleet's break rate more than halved.

  \end{abstract}
  \maketitle

  \section{Introduction}
\label{sec:intro}

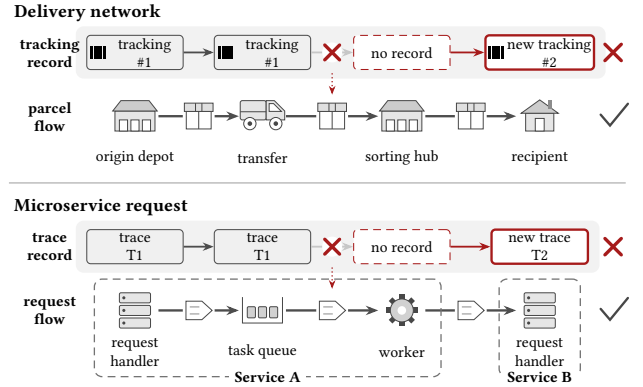
\begin{figure}[t]
  \centering
  \resizebox{0.98\columnwidth}{!}{%
    % ============================================================
% Figure 1 (single column) -- broken context propagation
% palette: black + one accent (cfAccent); safe in greyscale
% needs: figures/context/preamble.tex
% ============================================================
\begin{tikzpicture}[
  font=\sffamily, text=black,
  icn/.style={inner sep=0pt, minimum size=0.62cm, draw=none},
  rec/.style={draw=black!70, fill=black!6, rounded corners=1.5pt,
              minimum width=1.30cm, minimum height=0.46cm, inner sep=1pt, align=center,
              font=\tiny, text=black},
  norec/.style={rec, draw=cfAccent, fill=white, densely dashed},
  newrec/.style={rec, draw=cfAccent, line width=0.9pt, fill=white},
  work/.style={-{Stealth[length=1.7mm]}, line width=0.9pt, black!70},
  ctx/.style={-{Stealth[length=1.5mm]}, line width=0.7pt, black!70},
  guide/.style={line width=0.5pt, cfAccent, densely dotted},
  lane/.style={rounded corners=3pt},
  svc/.style={draw=black!55, densely dashed, rounded corners=3pt, line width=0.5pt},
  lab/.style={font=\tiny, text=black, inner sep=1pt, align=center},
  ttl/.style={font=\tiny\bfseries, scale=1.18, text=black, anchor=west, inner sep=1pt},
]
\newcommand{\hop}[3]{\draw[work] (#1) -- (#2);
  \fill[white] ($(#1)!0.5!(#2)$) circle (0.215);
  \pic[scale=0.86] at ($(#1)!0.5!(#2)$) {#3};}

\def\xa{0}\def\xb{1.72}\def\xc{3.60}\def\xd{5.46}
\def\xk{2.60}          % the handoff that drops the record
\def\xg{-1.15}          % lane-name gutter

% =========================== DELIVERY (analogy) =======================
\node[ttl] at (-1.68,6.12) {Delivery network};

\node[rec]    (t1) at (\xa,5.58) {\hspace{0.26cm}tracking\\\hspace{0.26cm}\#1};
\node[rec]    (t2) at (\xb,5.58) {\hspace{0.26cm}tracking\\\hspace{0.26cm}\#1};
\node[norec]  (t3) at (\xc,5.58) {no record};
\node[newrec] (t4) at (\xd,5.58) {\hspace{0.26cm}new tracking\\\hspace{0.26cm}\#2};
\bcode{t1}{black}\bcode{t2}{black}\bcode{t4}{black}

\draw[ctx] (t1) -- (t2);
\draw[ctx, black!22] (t2) -- (t3);
\draw[ctx, cfAccent] (t3) -- (t4);
\coordinate (kA) at ($(t2.east)!0.5!(t3.west)$);
\fill[white] (kA) circle (0.135); \pic[scale=0.95] at (kA) {ico cross};

\node[icn] (a1) at (\xa,4.72) {}; \pic at (a1) {ico depot};
\node[icn] (a2) at (\xb,4.72) {}; \pic at (a2) {ico truck};
\node[icn] (a3) at (\xc,4.72) {}; \pic at (a3) {ico depot};
\node[icn] (a4) at (\xd,4.72) {}; \pic at (a4) {ico home};
\foreach \i/\j in {1/2,2/3,3/4} {\hop{a\i}{a\j}{ico parcel}}

\node[lab] at (\xa,4.16) {origin depot};
\node[lab] at (\xb,4.16) {transfer};
\node[lab] at (\xc,4.16) {sorting hub};
\node[lab] at (\xd,4.16) {recipient};

\draw[guide, -{Stealth[length=1.2mm]}] ($(kA)+(0,-0.28)$) -- ($(a2)!0.5!(a3)+(0,0.29)$);

\node[lab, font=\tiny\bfseries] at (\xg,5.58) {tracking\\record};
\node[lab, font=\tiny\bfseries] at (\xg,4.72) {parcel\\flow};
\pic[scale=1.05] at (6.45,5.58) {ico cross};
\pic[scale=1.05] at (6.45,4.72) {ico check};

% =========================== separator ================================
\draw[black!30, line width=0.5pt] (-1.70,3.82) -- (6.62,3.82);

% =========================== MICROSERVICES ============================
\node[ttl] at (-1.68,3.50) {Microservice request};

\node[rec]    (s1) at (\xa,2.96) {trace\\T1};
\node[rec]    (s2) at (\xb,2.96) {trace\\T1};
\node[norec]  (s3) at (\xc,2.96) {no record};
\node[newrec] (s4) at (\xd,2.96) {new trace\\T2};

\draw[ctx] (s1) -- (s2);
\draw[ctx, black!22] (s2) -- (s3);
\draw[ctx, cfAccent] (s3) -- (s4);
\coordinate (kB) at ($(s2.east)!0.5!(s3.west)$);
\fill[white] (kB) circle (0.135); \pic[scale=0.95] at (kB) {ico cross};

\node[icn] (b1) at (\xa,2.10) {}; \pic at (b1) {ico server};
\node[icn] (b2) at (\xb,2.10) {}; \pic at (b2) {ico queue};
\node[icn] (b3) at (\xc,2.10) {}; \pic at (b3) {ico gear};
\node[icn] (b4) at (\xd,2.10) {}; \pic at (b4) {ico server};
\foreach \i/\j in {1/2,2/3,3/4} {\hop{b\i}{b\j}{ico req}}

\node[lab] (n1) at (\xa,1.54) {request\\handler};
\node[lab] (n2) at (\xb,1.54) {task queue};
\node[lab] (n3) at (\xc,1.54) {worker};
\node[lab] (n4) at (\xd,1.54) {request\\handler};

\draw[guide, -{Stealth[length=1.2mm]}] ($(kB)+(0,-0.28)$) -- ($(b2)!0.5!(b3)+(0,0.29)$);

\node[lab, font=\tiny\bfseries] at (\xg,2.96) {trace\\record};
\node[lab, font=\tiny\bfseries] at (\xg,2.10) {request\\flow};
\pic[scale=1.05] at (6.45,2.96) {ico cross};
\pic[scale=1.05] at (6.45,2.10) {ico check};

% service boundaries
\node[svc, fit=(b1)(b3)(n1)(n3), inner xsep=5pt, inner ysep=3pt] (svcA) {};
\node[svc, fit=(b4)(n4),         inner xsep=5pt, inner ysep=3pt] (svcB) {};
\node[lab, fill=white, inner xsep=2pt, font=\tiny\bfseries] at (svcA.south) {Service A};
\node[lab, fill=white, inner xsep=2pt, font=\tiny\bfseries] at (svcB.south) {Service B};

% =========================== lane shading =============================
\begin{scope}[on background layer]
  \node[lane, fill=black!5, fit=(t1)(t4), inner xsep=4pt, inner ysep=3pt] {};
  \node[lane, fill=black!5, fit=(s1)(s4), inner xsep=4pt, inner ysep=3pt] {};
\end{scope}

\end{tikzpicture}%
  }
  \caption{Broken context propagation and its analogy.}
  \Description{A delivery-network analogy above a microservice request. In
  both, the parcel or request crosses every handoff, but its tracking or trace
  record is lost at an intermediate handoff and restarted under a new
  identifier.}
  \label{fig:problem}
\end{figure}

A request entering a large microservice fleet rarely stays in one service. It fans out through queues, workers,
callbacks, and remote calls~\cite{gan2019deathstarbench}, and every downstream
operation must carry the request's context so its work can be traced,
attributed, and governed. When a handoff drops that context, work that should stay part of the
request continues without it. We call this \emph{broken context propagation}. The work still runs and the request succeeds, but the recorded
history fragments: the downstream call surfaces as a new root belonging to no
request. The asymmetry is familiar from parcel delivery: a depot that forwards
a package without scanning it still delivers, under a tracking record that
breaks midway (Figure~\ref{fig:problem}). In one snapshot of our own fleet, which
serves a major global short-video streaming platform, 673 of 1,133 services
carried at least one propagation break. Operators struggle to diagnose incidents, attribute
traffic, and plan capacity from a record that disagrees with the execution it
describes.

Shared infrastructure prevents breaks on recognized paths: tracing frameworks,
transport standards, and RPC middleware propagate context along the paths they
recognize~\cite{sigelman2010dapper,w3ctracecontext,opentelemetrycontext}. However, the
handoffs they do not recognize are hard to detect and hard to fix. The symptom is
ambiguous: a \emph{root-like downstream call} is equally consistent with a
genuine break and with periodic or deliberately detached work, and functional
tests pass in all of these cases. Copying the incoming context across
every handoff is no safer: a delayed retry must not inherit an expired
deadline, and a batch serving several requests has no single parent. Nor does
repairing the record suffice: the missing edges can be inferred offline from
telemetry~\cite{huye2024casper,ashok2024traceweaver}, but the same context
carries the deadline, identity, and priority labels that capacity decisions act
on, which must reach the running work. A repair
therefore has to begin at a production symptom that names no
source location and end at the handoff whose contract must change, in
application code, a dependency, or a shared library.

Our starting point is that a propagation break damages the record of an
execution without stopping the execution itself. The work still runs, so it can
be run again under controlled conditions, and what it does then tests whether a
break is real and whether a change repaired it. \sys{} is a specialized agentic
system that builds a repair loop on that surviving execution. Replaying the
target workload against an unmodified deployment and comparing it with an idle
baseline separates request-driven work from what a service generates on its
own. Source analysis then walks backward from the observed call through
application and dependency code, inspecting each handoff between execution
paths, until it reaches where the work continued but
its context did not. A bounded change restores the
causality, lifetime, and attribution the handoff requires, and the same replay,
run against the exact deployed revision, tests the result.

We have run this loop on our microservice fleet for 112 days. Across 26,136
calls in 2,587 services carrying 1.42 billion successful queries per second,
the fleet's actionable break rate more than halved (20.12\% to 8.55\%) and fell
by nearly a factor of four when weighted by traffic, while the monitored
service count grew 13.7\%. The decline held in 12 of 14 product domains and in
every implementation language. On a cohort of 240 calls in 16 services followed
through each repair's first production deployment, the break rate fell from
90.46\% to 4.69\%. On a controlled benchmark built from human-repaired breaks,
\sys{} reaches more targets for every coding agent and model we paired it
with, showing the contribution of its specialized Skills, tools, and knowledge
base beyond the agent and model.

Two lessons outlive context propagation. Automated repair normally starts from
something that fails: a crash, a failing test, a race report. Nothing fails
here, so the surviving execution becomes the test. A defect that corrupts what
a system records while leaving what it does intact, in metrics, audit trails,
or lineage, invites the same approach. The second lesson concerns fleets.
Frameworks and transport standards prevent breaks on the paths they recognize.
However, a fleet written in different languages by different teams always
contains paths the machinery misses. Having
an engineer read unfamiliar code at each missed path was never feasible, so
those paths accumulated into a long tail. An agentic system like \sys{}, equipped
with expert knowledge of the defect class and runtime validation, makes
repairing them one at a time affordable. It complements framework-level
prevention by repairing the residual long tail.

  \section{Background}
\label{sec:background}

\subsection{One Request, Many Services}

Large Internet applications can comprise hundreds or thousands of loosely
coupled microservices~\cite{gan2019deathstarbench}. Each service typically
implements a bounded function and can be developed, deployed, and scaled
independently. A user request, however, is rarely confined to one service. It
may trigger tens or hundreds of downstream operations before the
application assembles a response. Those operations can traverse services owned by different
teams, implemented in different languages, and changed on independent
schedules. Within a service, the request may also pass through asynchronous
tasks, queues, callbacks, workers, or shared libraries. These steps extend the
request beyond one process and often beyond one call stack.

Microservice decomposition enables decentralized development and independent
scaling, but it distributes each request's execution history across a large and
continuously changing fleet. No engineer has complete knowledge of every
participating service and execution path. Reconstructing what happened
requires the system to retain which request each downstream operation belongs
to.
Production systems therefore rely on \textbf{observability infrastructure} to record
what happened, in what order, and on behalf of which request. Engineers use this information to diagnose
incidents, recover service dependencies, attribute traffic and cost, plan
capacity, and place services. Observability is thus not
only a debugging aid. It is a principal means by which operators understand and
govern the system, so an incorrect execution history can lead to incorrect
operational conclusions even when the application runs as expected.

\subsection{How We Observe the System}

Distributed tracing is a standard mechanism for reconstructing a request's
execution across services. It plays a role analogous to a tracking number in a
delivery network: each package can pass through many independently operated
facilities, but the shared identifier allows its journey to be reconstructed.
Similarly, a traced request carries a globally unique \emph{trace identifier}.
Instrumentation records selected operations as \emph{spans}. Each non-root
span records a reference to its parent span. The spans connected by a trace
identifier and parent--child relationships form a \emph{trace}, the observed
history of the request.
Tracing systems such as X-Trace, Dapper, Zipkin, Jaeger, and
OpenTelemetry correlate spans across services by propagating \emph{trace
context}, which identifies the trace and the current parent
span~\cite{xtrace2007,sigelman2010dapper,twitterzipkin2012,uberjaeger2017,
opentelemetrytraces}. W3C Trace Context defines interoperable encodings for
carrying it across service boundaries~\cite{w3ctracecontext}.

Trace context is one part of the broader \emph{request context} associated with
an execution. Production systems may also propagate diagnostic state, routing
and traffic labels, test-traffic labels, identity and attribution data, or
policy and isolation metadata~\cite{mace2015pivot,w3cbaggage}. These fields
serve different purposes, but all can lose their meaning if they become
detached from the request to which they belong.

\subsection{Context Propagation}

The transfer of this metadata along execution paths is called \emph{context
propagation}~\cite{mace2018universal,opentelemetrycontext}. Within one process,
context is a value associated with the current execution. A program may pass
this value explicitly between functions, or a language runtime may make it
available implicitly through execution-local state. Later calls read the
trace identifier, current span, and other request-scoped values from this
context. Context must also cross local execution handoffs: a program that
submits request-driven work to a queue must arrange for the worker to receive
that context when it executes the task.

An in-memory context value is local to one process. When a remote procedure
call (RPC) crosses a process boundary, the sender serializes the relevant fields
into HTTP headers or RPC metadata~\cite{opentelemetrypropagators}. The receiver
extracts those fields and reconstructs a local context. Tracing instrumentation
can then create a child span, and subsequent local or remote operations can
continue the trace. Context therefore travels as a program value within a
process and as encoded fields between processes. Shared RPC frameworks automate
cross-process propagation when execution remains within the boundaries
they understand.

\textbf{Correct propagation maintains the request's causal chain,} the
operations causally related to the original
request~\cite{xtrace2007,sambasivan2016principled}.
Every downstream operation that \emph{remains part of the request} must be
associated with it, while \emph{unrelated background work} must remain
outside. Request context carries this association across handoffs.
\S\ref{sec:problem} examines why this requirement becomes
difficult when execution crosses handoffs outside automatic instrumentation.

  \section{Broken Context Propagation at Scale}
\label{sec:problem}

\subsection{Failure Semantics and Symptoms}
\label{sec:failure-semantics}

A request entering a service may produce new operations through asynchronous
execution, queues, batching, retries, and shared libraries. At each such
handoff, the system must determine whether the resulting operation still
belongs to the original request and preserve the causal
chain. We study this application-level causal chain. A \emph{context
propagation break}, or propagation break, occurs when work that should remain
associated with the request continues without the context needed to
preserve that association.

Figure~\ref{fig:problem} shows a common example. A handler enqueues a task's
business payload, moving execution to a worker. Unless the payload also carries
the context, the worker starts without it: neither the context value the
handler passed explicitly nor the execution-local state its runtime holds
crosses the queue. The worker can still execute the task
and issue the RPC, but trace $T_1$ is no longer available. Client instrumentation
may therefore begin a new trace $T_2$, making the break at the queue visible
only at the later RPC.

The RPC under $T_2$ has no visible request parent, which we call a
\emph{root-like downstream call}; $T_2$ itself is a disconnected trace.
The root-like call is a diagnostic candidate, not proof of a
defect: periodic, startup, maintenance, and deliberately detached work may
legitimately begin as roots. Outbound call metrics derive the caller
method from propagated context; without that context, they report the call
as root-like. Propagation breaks and legitimate root-like
calls therefore look the same. Diagnosis must determine whether the
downstream work was request-correlated.

Passing the incoming context across every handoff is not a safe repair. Correct
propagation answers three questions: \emph{causality}, whether later work belongs
to the request's causal chain; \emph{lifetime}, whether it may outlive the
request's deadline or cancellation scope; and \emph{attribution}, whether it
belongs to one request, several requests, or none. Parallel work may inherit
context directly; a delayed retry may retain ancestry without an expired
deadline; a multi-request batch has no unique parent. A repair must preserve the
appropriate causality, lifetime, and attribution at each handoff rather than
propagate context everywhere.
Recovering the chain offline fails at the other end: it can restore causality in
the record, but lifetime and attribution constrain what the execution does and
must hold while it runs.

Propagation breaks are not specific to one language or context API. Go passes
\texttt{context.Context} explicitly, whereas OpenTelemetry Java propagates
thread-local context through executor wrappers~\cite{gocontext,
oteljavacontext}. Python and Node.js provide analogous mechanisms for managed
asynchronous work~\cite{pythonasyncio,nodeasynccontext}.
The mechanisms differ, but each propagates context only along
execution paths they recognize. A handoff outside those paths requires an
explicit decision about whether and how the request's causal chain continues.

\subsection{Why Missing Context Matters}

A request can succeed after a propagation break while observability and control
systems receive an incomplete or incorrect account of its execution.

\paragraphb{During incidents.} A trace can omit real dependency edges, make a
downstream operation appear to be an independent root, and deprive automated
diagnosis of relevant evidence. Analyses that begin from a visible request can
silently exclude the detached fragment. Responders must then reconstruct the
missing causal chain from weaker signals such as logs, metrics, and source
code.

\paragraphb{During normal operation.} Operators aggregate traces into service
dependency graphs and attribute downstream traffic to its request
origin~\cite{sigelman2010dapper}. Capacity planning and placement depend on
knowing which upstream workloads create demand at each service~\cite{orion2022}. Routing and
priority mechanisms similarly use request origin and labels to decide how work
should be handled~\cite{colarusso2024locality}. A missing edge can assign work
to the wrong origin, hide a communication relationship, or drop diagnostic,
identity, test-traffic, isolation, or policy metadata carried with the request.

In both settings, we act on a record that
disagrees with the execution it describes.
\textbf{The desired result is a trustworthy causal chain.}
It preserves the call topology, attributes latency and errors to the
correct request, excludes unrelated downstream work, and respects the
application's concurrency, timeout, and lifecycle semantics. Observability can be
corrected afterward; control decisions depend on live requests carrying the
correct context.

\subsection{Production Measurement}
\label{sec:production-measurement}

Four measurements from a large production microservice fleet show the
problem's breadth, range of diagnostic outcomes, distance from symptoms to
fixes, and manual cost.

\begin{figure}[t]
  \centering
  \captionsetup{skip=2pt}
  \captionsetup[subfigure]{skip=1pt}
  \begin{subfigure}[b]{0.4\columnwidth}
    \centering
    \includegraphics[width=\linewidth]{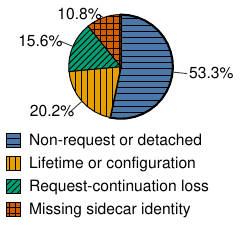}
    \caption{Diagnostic outcomes.}
    \label{fig:problem-outcomes}
  \end{subfigure}\hfill
  \begin{subfigure}[b]{0.4\columnwidth}
    \centering
    \includegraphics[width=\linewidth]{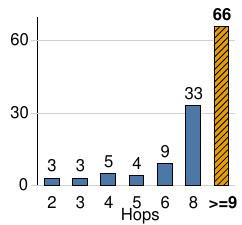}
    \caption{Repair distance.}
    \label{fig:problem-depth}
  \end{subfigure}
  \caption{Production outcomes and repair distances.}
  \Description{A pie chart shows four diagnostic outcome classes among 1,088
  labeled root-like calls. A compact vertical bar chart shows the service-graph
  distance from 123 observed symptoms to their repair sites.}
  \label{fig:problem-measurements}
\end{figure}

\paragraphb{The problem is widespread.} In a snapshot taken on July 28, 2025,
673 of 1,133 services (59.4\%) were classified as having at least one
propagation break. We found the candidates in the outbound call metrics, which
report a call as root-like when the caller-method field contains the callee's
own service rather than an upstream method. For the subset of services with mesh support,
we also injected a test header at ingress and checked whether it reached egress.

\paragraphb{The same symptom requires different responses.} We inspected
1,088 root-like calls from 73 caller services and labeled them into four classes. As Figure~\ref{fig:problem-outcomes} shows, 53.3\% were
not caused by a user request or were intentionally detached, while 15.6\% were
request continuations that had lost context. Lifecycle or configuration
mismatches accounted for 20.2\%, and calls attributed to sidecars without an
independent service identity accounted for 10.8\%. These outcomes reflect
different underlying questions: whether the call was request-driven, whether
its execution lifetime matched the available context, and whether the observed
caller represented an independent service. The same root-like call can
therefore require no change, an application repair, a configuration change, or
a change to how the caller is represented.

\paragraphb{Symptoms are far from their repair sites.} For 123
SDK-linked broken-call samples, we measured the maximum observable upstream
service-graph depth. Of these samples, 91.1\% were at depth 5 or greater,
and 80.5\% were at depth 8 or greater (Figure~\ref{fig:problem-depth}). The
observed broken call is therefore often far downstream of the shared propagation
mechanism that must be understood or repaired, making the defect harder to fix.

\paragraphb{Manual diagnosis is costly.} Across a 20-case operational
baseline, the median repaired break required 90 minutes even before review and
rollout. At the measured fleet scale, new breaks can accumulate faster than
engineers can manually clear them, making repair a recurring backlog rather
than a one-time cleanup.

\subsection{Why Diagnosis and Repair Are Difficult}
\label{sec:repair-challenges}

Broken propagation is a structural consequence of microservice architecture,
not simply careless implementation. Microservice decomposition makes context
propagation a global invariant implemented by many locally owned handoffs, with
no component that owns the complete path~\cite{gan2019deathstarbench}.

\paragraphb{The failed handoff and its owner are not known in advance.} Shared RPC and
HTTP frameworks control standard request entry and client exit points, but real
services also use custom executors, queues, callbacks, batching layers, and
shared libraries. Even within Go, these handoffs take heterogeneous forms:
goroutines, queues, custom wrappers, and indirect exchanges through shared
memory. These forms resist fixed string or local AST patterns. A break can occur at
any of them, including within a single service. The business payload may still
reach the downstream service, and the RPC may succeed, so
request-success monitoring does not reveal that its context was lost.
A large, rapidly evolving fleet also spans product domains, languages, framework
versions, legacy components, coding conventions, and ownership boundaries. The
repair may therefore belong in application code, a dependency, or a shared
library; a repair method must discover both the failed handoff and its owner.

\paragraphb{Failures are silent and reveal only downstream symptoms.} Work
continues after its context is lost, so compilation and functional tests can
pass while the recorded execution history is wrong. Monitoring may reveal a
new trace root at an RPC client, but not whether the context disappeared in a
business function, dependency library, queue submission, or worker execution.
The missing edge also obscures the runtime path needed to localize the failure;
as \S\ref{sec:production-measurement} shows, the observed symptom can be far
from the responsible handoff. Source code describes which paths are possible,
but cannot establish which path a particular production request followed. Even
after a candidate handoff is found, a repair that compiles or passes functional
tests does not establish that the worker received the correct context or that
the outgoing RPC used the intended parent. Runtime evidence is therefore
required both to establish the break and to test the deployed repair.

\paragraphb{No single rule fits every handoff.} One request
may create several asynchronous tasks, while a batch may combine work from
several requests. A retry may execute after the original request has ended, and
a timer or initialization task may belong to no request. Context fields also
have different lifetimes: preserving trace ancestry does not necessarily
justify inheriting an expired deadline, cancellation signal, identity, or
policy label. Correct repair must answer the
\emph{causality}, \emph{lifetime}, and \emph{attribution} questions from
\S\ref{sec:failure-semantics}. A uniform rule that
copies the incoming context can invent a false parent, misattribute shared
work, or impose a request's deadline on work with a different
lifetime.

\subsection{Limits of Existing Approaches}
\label{sec:existing-limits}

Existing prevention mechanisms get the central principle right: context
propagation is too fragile to leave to each application, so shared
infrastructure should carry context automatically across the boundaries it
controls~\cite{xtrace2007,sigelman2010dapper,sambasivan2016principled,
kaldor2017canopy,mace2018universal,w3ctracecontext,opentelemetrycontext}. In
our Go stack, CloudWeGo LocalSession held incoming request context in
goroutine-local storage (GLS), and Kitex RPC and Hertz HTTP middleware
recovered it before outgoing
RPCs~\cite{cloudwegolocalsession,cloudwegokitex,cloudwegohertz}.

Prevention operates at complementary layers. Transport standards define
how context is encoded across
services~\cite{w3ctracecontext,w3cbaggage}. Context libraries
provide propagation abstractions~\cite{mace2018universal,opentelemetrycontext}
and support inject and extract operations at process
boundaries~\cite{opentelemetrypropagators}. Shared libraries and framework
middleware automate recognized execution
paths~\cite{sigelman2010dapper,cloudwegokitex,cloudwegohertz}. Their explicit
propagation and fallback recovery both assume that the relevant handoff is
recognized or instrumented; otherwise, request-scoped state remains on the original execution
path. These mechanisms cannot discover an unknown queue, callback, batch, or
library handoff, recover context lost before an RPC, or decide whether the
resulting work should inherit the request.

Once the intended propagation path is known, source transformation can automate
mechanical edits. Prior work on automated context propagation for Go rewrites
function signatures and call sites from configured use sites~\cite{welc2021contextpropagation}.
Static checks can flag known source patterns, while trace or log analysis can
reveal missing structure. These tools reduce editing or supply evidence, but
they do not decide whether the symptom is a defect, where context was first
lost, what should cross the handoff, which codebase should change, or whether the
deployed repair restored the causal chain.

Our work complements this prevention with a production-guided loop for residual
symptoms (\S\ref{sec:design}). It determines whether a symptom is a defect,
locates the uncovered handoff, restores the required causality, lifetime, and
attribution, and validates the deployed result. Because software evolution
keeps producing new breaks, repair proceeds in bounded waves across the
microservice fleet.

  \section{\sys{} Design}
\label{sec:design}

\begin{figure*}[t]
  \centering
  \resizebox{0.95\textwidth}{!}{%
    % ============================================================
% Figure 2 (double column, flat) -- ctx-fixer design
% palette: black + one accent (cfAccent); safe in greyscale
% needs: figures/context/preamble.tex
% ============================================================
\begin{tikzpicture}[
  font=\sffamily, text=black,
  icn/.style={inner sep=0pt, minimum size=0.62cm, draw=none},
  rec/.style={draw=black!70, fill=black!6, rounded corners=1.5pt,
              minimum width=1.20cm, minimum height=0.36cm, inner sep=1pt,
              align=center, font=\tiny, text=black},
  norec/.style={rec, draw=cfAccent, fill=white, densely dashed, minimum width=1.05cm},
  newrec/.style={rec, draw=cfAccent, line width=0.9pt, fill=white},
  outc/.style={rounded corners=1.5pt, minimum width=1.05cm, minimum height=0.32cm,
               inner sep=1pt, font=\tiny, text=black},
  work/.style={-{Stealth[length=1.7mm]}, line width=0.9pt, black!70},
  ctx/.style={-{Stealth[length=1.5mm]}, line width=0.7pt, black!70},
  sflow/.style={-{Stealth[length=1.4mm]}, line width=0.65pt, black!70},
  guide/.style={line width=0.5pt, cfAccent, densely dotted},
  lane/.style={rounded corners=3pt},
  ring/.style={circle, draw=black!70, fill=white, minimum size=0.86cm, inner sep=0pt},
  loop/.style={-{Stealth[length=1.9mm]}, line width=1.0pt, black!70},
  kb/.style={-{Stealth[length=1.6mm]}, line width=0.8pt, black!70, densely dashed},
  panel/.style={rounded corners=4pt, draw=black!25, fill=black!2},
  replaybox/.style={rounded corners=2pt, draw=black!38, fill=white,
                    line width=0.5pt},
  lab/.style={font=\tiny, text=black, inner sep=1pt, align=center},
  blab/.style={lab, font=\tiny\bfseries},
  hdr/.style={blab, text=black!75},
  ttl/.style={font=\tiny\bfseries, scale=1.18, text=black, anchor=west, inner sep=1pt},
]
\newcommand{\hop}[3]{\draw[work] (#1) -- (#2);
  \fill[white] ($(#1)!0.5!(#2)$) circle (0.215);
  \pic[scale=0.86] at ($(#1)!0.5!(#2)$) {#3};}

% =====================================================================
% (a) two planes under one controlled workload
% =====================================================================
\node[ttl] at (-1.58,5.30) {(a) Two planes, one workload};

\def\ya{0}\def\yb{1.55}\def\yc{3.10}\def\yd{4.82}
\def\xg{-1.05}

\node[rec]    (v1) at (\ya,4.58) {Trace T1};
\node[rec]    (v2) at (\yb,4.58) {Trace T1};
\node[norec]  (v3) at (\yc,4.58) {No record};
\node[newrec] (v4) at (\yd,4.58) {New trace T2};
\draw[ctx] (v1) -- (v2);
\draw[ctx, black!22] (v2) -- (v3);
\draw[ctx, cfAccent] (v3) -- (v4);
\coordinate (kk) at ($(v2.east)!0.5!(v3.west)$);
\fill[white] (kk) circle (0.135); \pic[scale=0.95] at (kk) {ico cross};

\node[icn] (x1) at (\ya,3.54) {}; \pic at (x1) {ico server};
\node[icn] (x2) at (\yb,3.54) {}; \pic at (x2) {ico queue};
\node[icn] (x3) at (\yc,3.54) {}; \pic at (x3) {ico gear};
\node[icn] (x4) at (\yd,3.54) {}; \pic at (x4) {ico server};
\foreach \i/\j in {1/2,2/3,3/4} {\hop{x\i}{x\j}{ico req}}
\draw[guide, -{Stealth[length=1.2mm]}]
      ($(kk)+(0,-0.28)$) -- ($(x2)!0.5!(x3)+(0,0.29)$);

\node[blab] (m1) at (\ya,3.02) {Handler};
\node[blab] (m2) at (\yb,3.02) {Task queue};
\node[blab] (m3) at (\yc,3.02) {Worker};
\node[blab] (m4) at (\yd,3.02) {Service B};

\node[blab] (traceLabel) at (\xg,4.58) {Trace\\record};
\node[blab] (requestLabel) at (\xg,3.54) {Request\\flow};
\node[icn] (mk1) at (5.80,4.58) {}; \pic[scale=1.05] at (mk1) {ico cross};
\node[icn] (mk2) at (5.80,3.54) {}; \pic[scale=1.05] at (mk2) {ico check};

\draw[black!22, line width=0.5pt] (-1.38,2.81) -- (5.88,2.81);

% ---- replay the workload and read off the outcome ----
\node[hdr] (hd1) at (0.30,2.60) {Replayed traffic};
\node[hdr] (hd2) at (2.05,2.60) {Target service};
\node[hdr] (hd3) at (3.95,2.60) {Broken trace};

\node[icn] (g1) at (0.30,2.10) {}; \pic[scale=0.86] at (g1) {ico flat};
\node[icn] (g2) at (0.30,1.58) {}; \pic[scale=0.86] at (g2) {ico pulse};
\node[icn] (t1) at (2.05,2.10) {}; \pic[scale=0.86] at (t1) {ico server};
\node[icn] (t2) at (2.05,1.58) {}; \pic[scale=0.86] at (t2) {ico server};
\node[outc, draw=cfAccent, densely dashed, fill=white] (o1) at (3.95,2.10) {Absent};
\node[outc, draw=cfAccent, line width=0.9pt, fill=black!12] (o2) at (3.95,1.58) {Present};
\foreach \i in {1,2} {\draw[sflow] (g\i) -- (t\i); \draw[sflow] (t\i) -- (o\i);}
\node[blab, anchor=east] (idleLabel) at (-0.12,2.10) {Idle};
\node[blab, anchor=east] (loadLabel) at (-0.12,1.58) {Load};

% Align the controlled replay box with the separator above it.
\coordinate (replayTop) at (2.00,2.81);
\node[fit=(replayTop)(idleLabel)(loadLabel)(hd1)(hd2)(hd3)
          (g1)(g2)(t1)(t2)(o1)(o2),
      inner xsep=4pt, inner ysep=0pt] (replayBox) {};

\node[lab] (concl) at (2.20,0.98)
  {The orphan tracks the request traffic, so this work is owed a parent};

% =====================================================================
% (b) local repair loop
% =====================================================================
\node[ttl] at (6.86,5.30) {(b) Local repair loop};
\def\lx{8.78}\def\ly{3.10}\def\lr{1.45}

\node[icn] at (\lx,\ly) {}; \pic at (\lx,\ly) {ico server};
\node[blab] at (\lx,\ly-0.48) {One service};

\node[ring] (p1) at ($(\lx,\ly)+(135:\lr)$) {}; \pic at (p1) {ico glass};
\node[ring] (p2) at ($(\lx,\ly)+(45:\lr)$)  {}; \pic at (p2) {ico target};
\node[ring] (p3) at ($(\lx,\ly)+(-45:\lr)$) {}; \pic at (p3) {ico wrench};
\node[ring] (p4) at ($(\lx,\ly)+(-135:\lr)$){}; \pic at (p4) {ico scale};

\draw[loop] (p1) to[bend left=22] (p2);
\draw[loop] (p2) to[bend left=22] (p3);
\draw[loop] (p3) to[bend left=22] (p4);
\draw[loop] (p4) to[bend left=22] (p1);

\node[blab, above=2pt of p1] (pReplay) {Replay};
\node[blab, above=2pt of p2] (pLocalize) {Localize};

% =====================================================================
% (c) fleet repair loop
% =====================================================================
\node[ttl] at (11.62,5.30) {(c) Fleet repair loop};
\def\fx{13.55}\def\fy{3.10}

\node[icn] at (\fx,\fy) {}; \pic[scale=1.15] at (\fx,\fy) {ico fleet};
\node[blab] at (\fx,\fy-0.52) {Visible\\repair set};

\node[ring] (q1) at ($(\fx,\fy)+(135:\lr)$) {}; \pic at (q1) {ico chart};
\node[ring] (q2) at ($(\fx,\fy)+(45:\lr)$)  {}; \pic at (q2) {ico funnel};
\node[ring] (q3) at ($(\fx,\fy)+(-45:\lr)$) {}; \pic at (q3) {ico wave};
\node[ring] (q4) at ($(\fx,\fy)+(-135:\lr)$){}; \pic at (q4) {ico eye};

\draw[loop] (q1) to[bend left=22] (q2);
\draw[loop] (q2) to[bend left=22] (q3);
\draw[loop] (q3) to[bend left=22] (q4);
\draw[loop] (q4) to[bend left=22] (q1);

\node[blab, above=2pt of q1] (qMeasure) {Measure};
\node[blab, above=2pt of q2] (qSelect) {Select};

% ---- loop labels ----
\node[blab, fill=white, inner sep=1.5pt, below=2pt of p3] (lb1) {Repair};
\node[blab, fill=white, inner sep=1.5pt, below=2pt of p4] (lb2) {Validate};
\node[blab, fill=white, inner sep=1.5pt, below=2pt of q3] (lb3) {Repair wave};
\node[blab, fill=white, inner sep=1.5pt, below=2pt of q4] (lb4) {Re-observe};

% Shared vertical anchors keep the local and fleet panels exactly aligned.
\coordinate (localTop) at (\lx,4.54);
\coordinate (localBottom) at (\lx,1.32);
\coordinate (fleetTop) at (\fx,4.54);
\coordinate (fleetBottom) at (\fx,1.32);
\begin{scope}[on background layer]
  \node[panel,
        fit=(p1)(p2)(p3)(p4)(pReplay)(pLocalize)(lb1)(lb2)(localTop)(localBottom),
        inner xsep=9pt, inner ysep=0pt] (localPanel) {};
  \node[panel,
        fit=(q1)(q2)(q3)(q4)(qMeasure)(qSelect)(lb3)(lb4)(fleetTop)(fleetBottom),
        inner xsep=9pt, inner ysep=0pt] (fleetPanel) {};
\end{scope}

% =====================================================================
% couplings and the shared knowledge base
% =====================================================================
\draw[work] (6.40,\ly) -- (7.04,\ly);
\node[blab, fill=white, inner sep=1.5pt] at (6.72,\ly-0.28) {Orphan};
\draw[kb] ($(p4.west)+(-0.03,0)$) -- (replayBox.east |- p4.west);
\node[blab, fill=white, inner sep=1.5pt]
  at ($(p4.west)!0.5!(replayBox.east |- p4.west)+(0,0.27)$) {Same workload};

% The agent mediates promotion of a validated local fix to the fleet loop.
% Derive its center from the actual facing panel boundaries.
\coordinate (agentCenter) at
  ($(localPanel.north east)!0.5!(fleetPanel.north west)+(0,-0.67)$);
\draw[work] (localPanel.east) -- (fleetPanel.west);
\node[circle, draw=black!40, fill=white, line width=0.5pt,
      minimum size=1.34cm, inner sep=0pt] (agent) at (agentCenter) {};
\pic[scale=1.25] at ($(agent)+(0,0.16)$) {ico bot};
\node[blab, inner sep=1pt] at ($(agent)+(0,-0.36)$) {Agent};
\node[blab, fill=white, inner sep=1pt]
  at ($(localPanel.east)!0.5!(fleetPanel.west)+(0,-0.28)$) {Proven fix};

\node[icn] (bk) at (11.16,1.18) {}; \pic[scale=1.0] at (bk) {ico book};
\node[blab] at (11.16,0.66) {Repair knowledge base};
\draw[kb] ($(q4.west)+(-0.02,-0.10)$) to[out=200,in=60] ($(bk)+(0.20,0.22)$);
\draw[kb] ($(bk)+(-0.22,0.20)$) to[out=125,in=-25] ($(p3.east)+(0.02,-0.10)$);

% =====================================================================
% remaining backgrounds
% =====================================================================
\begin{scope}[on background layer]
  \node[lane, fill=black!5, fit=(v1)(v4), inner xsep=4pt, inner ysep=3pt] {};
  \node[panel, fit=(traceLabel)(requestLabel)(v1)(v4)(x1)(x4)(m1)(m4)(replayBox)(concl)(mk1)(mk2),
        inner xsep=7pt, inner ysep=3.5pt] {};
  \node[replaybox, fit=(replayBox), inner sep=0pt] {};
\end{scope}

\end{tikzpicture}%
  }
  \caption{\sys{} design.}
  \Description{The design has three parts: controlled workload replay compares
  request flow with trace records; a per-service loop replays, localizes,
  repairs, and validates a break; and a fleet loop measures, selects, repairs,
  and re-observes bounded waves while sharing validated repair patterns.}
  \label{fig:design}
\end{figure*}

\subsection{Design Overview: Execution as the Reference}

The central observation behind \sys{} is that a propagation break damages the
record of an execution without necessarily damaging the execution itself. In
the delivery network of Figure~\ref{fig:problem}, a facility that forwards a
package without associating the outgoing handoff with its tracking number still
delivers it. The package's continued movement is then independent evidence that
the handoff occurred despite the missing record.

Distributed applications have the same two planes. The \emph{execution plane}
contains the handlers, tasks, queues, workers, and RPCs that perform a request.
It corresponds to the package and the facilities that move it. The
\emph{observation plane} holds the recorded spans and parent--child
relations that explain why each operation occurred. It corresponds to the delivery
tracking record formed by the tracking number and scans. At an uncovered handoff, the business payload can continue through the
execution plane while the observation plane records the resulting work as a new root.
As \S\ref{sec:existing-limits} explains, RPC and HTTP frameworks propagate
context across the request paths they recognize. Custom queues, callbacks,
workers, and library handoffs can move work to another path while transferring
only its payload, leaving the context behind.
\sys{} diagnoses these uncovered handoffs and restores the causality, lifetime,
and attribution each case requires, while retaining framework-level propagation
on the paths it already protects.

The key idea is to use the surviving execution as the reference throughout
repair. Because the target work still runs, \sys{} can replay it, follow its
execution path through handoffs, and validate the repaired causal chain against
the same execution. Controlled replay is the test for this failure class:
functional tests confirm that the business result succeeds, but say nothing
about the observation plane that the break damages. Together with source
analysis, replay establishes whether the call is
request-correlated and where execution first continued without its context.
The repair then restores the invariant from
\S\ref{sec:failure-semantics}: preserve the request's causal chain for work that
belongs to it, without attaching unrelated work.

\paragraphb{Repairing one break.} The local loop in
Figure~\ref{fig:design}(b) has four stages. (1) To determine whether a real
parent--child edge is missing, controlled traffic replay compares an idle
baseline with target-request traffic and tests whether the root-like call changes, distinguishing
request-driven work from independent activity. (2) Source analysis enumerates
the paths that can produce
the call and finds the earliest supported handoff where work continued but
context did not. (3) Evidence from the surrounding path determines
the required causality, lifetime, and attribution, after which \sys{} changes
the smallest boundary that both has the correct context and controls the
receiving execution. For example, a request-scoped task may require context to
be captured at submission, restored only during execution, and cleared before
the worker accepts another task. (4) The system repeats the intervention to
verify that the same work still executes, its causal chain is restored, and
unrelated work does not inherit its context.

\paragraphb{Repairing the fleet.} A fleet cannot be treated as a fixed backlog
because a damaged observation plane may hide downstream breaks, while software
evolution creates new ones. Figure~\ref{fig:design}(c) therefore repeats the
local loop over the currently visible repair set: measure, repair a bounded
wave, and re-observe. Engineers promote recurring validated repairs into the
knowledge base, so later cases reuse patterns that a human approved.

\paragraphb{Why the repair is feasible now.} An engineer could always repair
one break; what did not scale was repeating the investigation required for it across
unfamiliar services. That investigation also resists the fixed string and AST
rules that cover only anticipated syntax (\S\ref{sec:repair-challenges}).
What changed is that
tool-using large language model (LLM) agents now combine broad code
understanding with iterative tool use: they form a path hypothesis, search
multiple repositories, retrieve missing evidence, and revise it from source and
runtime feedback~\cite{yao2023react,yang2024sweagent}. \sys{} uses this
adaptability to propose bounded, case-specific changes, and
\S\ref{sec:design-agent} explains why the surrounding system rather than the
agent makes them acceptable.

\subsection{Diagnosing and Repairing One Break}
\label{sec:design-one-break}

Each stage of the local loop answers a challenge from
\S\ref{sec:repair-challenges}. The symptom is silent, so runtime evidence must
confirm that the break is real and later test whether the deployed repair
removed it. Neither the handoff nor its owner is known in advance, so source
analysis must find both. No single rule fits every handoff, so the repair is
chosen case by case.

\subsubsection{Diagnose}
\label{sec:design-diagnose}

Diagnosis begins with a suspicious root-like downstream call reported by
production monitoring. Monitoring aggregates these observations by caller
service, callee service, and callee method. We call this three-field key a
\emph{call tuple}. As \S\ref{sec:failure-semantics} established, a root-like
call is not necessarily a break, so \sys{} first determines whether it is one
and then localizes it in source code.

\paragraphb{Replay: Determine whether a parent--child edge is missing.} The system first
compares two conditions on the unchanged service: a baseline without the target
request traffic and an intervention that replays the target workload. A
root-like call that appears only after the intervention is evidence of
request-correlated work. A call already present in the baseline implies initialization, timers, maintenance, or other self-generated
work. The replayed workload becomes a persistent test case for later
validation. This comparison establishes whether the call is request-correlated, but source evidence must still determine whether the detachment is a defect and
which execution path caused it.

\paragraphb{Localize: Search by execution handoff.} Framework-managed request entry and
RPC exit normally preserve context; the uncontrolled portion is the execution
path between them. Starting from every callsite for the target RPC, \sys{} walks
backward through application and dependency source to each recognizable origin.
Rather than treating every function as a candidate, it inspects the handoffs
where work moves between execution paths, including queues, workers, callbacks,
channels, and library interfaces. On each discoverable path, it finds the
last point that holds the correct context and the first representation of the
work that no longer carries it. This boundary is the \emph{repair anchor}. In
the queue example from \S\ref{sec:failure-semantics}, the handler still has
$T_1$, but the queued task does not, so the submit--run handoff is the anchor.
When runtime evidence cannot distinguish among multiple plausible paths, \sys{} analyzes
each path and records unresolved edges before selecting a repair.

\subsubsection{Fix: Restore the Handoff Contract}
\label{sec:design-repair}

Locating a boundary does not determine what context should cross it. As
\S\ref{sec:failure-semantics} established, the repair depends on causality,
lifetime, and attribution. \sys{} therefore preserves only the context
justified by these properties for each path.

\begin{figure}[t]
  \centering
  \resizebox{0.85\columnwidth}{!}{%
    % Compact repair-contract taxonomy for a single column.
% Needs: figures/context/preamble.tex
\begin{tikzpicture}[
  font=\sffamily\scriptsize,
  contract/.style={draw=cfBlue!75!black, fill=cfBlue!8, rounded corners=2pt,
    line width=0.5pt, text width=2.05cm, minimum height=0.66cm,
    align=center, font=\sffamily\scriptsize\bfseries, inner sep=2pt},
  semantics/.style={draw=black!45, fill=white, rounded corners=2pt,
    line width=0.45pt, text width=4.76cm, minimum height=0.66cm,
    align=left, inner sep=2.5pt},
  relation/.style={-{Stealth[length=1.3mm]}, line width=0.6pt, black!60},
]

\foreach \y/\title/\body [count=\row] in {
  0.00/{Create a\\legitimate\\root}/{Assign periodic and fan-in work a labeled root; repair initialization only when recurring noise justifies it.},
  -0.83/{Preserve request\\causality}/{Carry context across the handoff, restore it for the corresponding work, then clear it.},
  -1.66/{Restore the\\framework span}/{Bind an existing framework entry span to worker execution without creating a new root.},
  -2.56/{Correct async\\recording}/{Configure or upgrade tracing so an asynchronous child remains attached after its parent finishes.},
  -3.39/{Separate platform\\traffic}/{Tag sidecar calls so measurement can exclude them without application changes.}
}{
  \node[contract] (contract-\row) at (1.05,\y) {\title};
  \node[semantics, anchor=west] (semantics-\row) at (2.42,\y) {\body};
  \draw[relation] (contract-\row.east) -- (semantics-\row.west);
}

\end{tikzpicture}%
  }
  \caption{Repair families distinguish the handoff contracts.}
  \Description{Five repair contracts map to their repair semantics: creating a
  legitimate root, preserving request causality, restoring a framework span,
  correcting asynchronous recording, and separating platform traffic.}
  \label{fig:repair-contracts}
\end{figure}
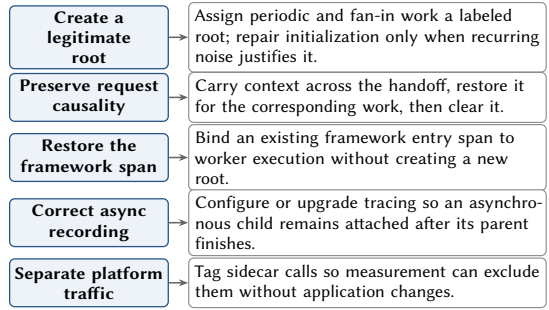

Figure~\ref{fig:repair-contracts} groups the resulting repairs by the contract
they restore. These families separate semantically different cases that can
produce the same root-like symptom. The knowledge base refines each family into
recognition patterns and implementation-specific actions
(\S\ref{sec:implementation-repair-validation}), but the semantic decision
remains stable across languages and frameworks.
The repair families enforce causality,
lifetime, and attribution. In Figure~\ref{fig:problem}, causality requires the
handler to attach $T_1$'s metadata to its queued task. Lifetime requires
preserving ancestry without inheriting the handler's deadline or cancellation.
Attribution permits the worker to restore $T_1$ because the task belongs to one
request, but requires clearing it afterward.

\sys{} localizes the smallest sufficient repair to the module that owns the
violated handoff contract. This follows information hiding and design for
extension: confining a design decision to its owning module limits affected
components and avoids duplicated workarounds~\cite{parnas1972criteria,
parnas1979extension}. Thus, a shared executor is repaired centrally, whereas
caller-specific semantics remain in application code.

\subsubsection{Validation}
\label{sec:design-validation}

Functional tests cannot establish that context propagation is correct because
the work may succeed despite a broken causal record. Every proposed repair must
therefore be validated at runtime. Closing the loop with
\S\ref{sec:design-diagnose}, \sys{} deploys the repaired version, injects the
same controlled traffic used for diagnosis, and compares its metrics with the
original production baseline.

A repair passes validation only if (1) the workload still reaches the target call,
(2) normal tests pass, (3) the root-like observation
at the target call tuple disappears under the same traffic, and (4) no
unrelated work has begun to inherit the request's context. Otherwise, the repair is rejected.

\subsection{Fleet-Wide Repair}
\label{sec:design-fleet}

Fleet scale changes orchestration, not the repair procedure. \sys{} remains a
per-service repair engine that diagnoses, repairs, and validates one service at
a time. A separate campaign layer prioritizes services, schedules bounded
batches, persists progress, handles retries, and measures aggregate progress
across these independent runs.

\paragraphb{Coordinate independent service repairs.}
The campaign ranks services by operational impact and materializes one work
item per service. Each item has its own workspace and checkpoints, so a failed
or paused repair can resume without affecting others. It runs only a
bounded number of items concurrently and records whether each item is
queued, running, validated, retryable, blocked, or awaiting review.

\paragraphb{Repeatedly update the visible repair set.}
Fleet monitoring aggregates outbound calls into the call tuples defined above.
Each tuple with a root-like observation becomes a \emph{candidate call tuple},
or candidate, in the visible repair set. The campaign filters for real breaks
and groups them by caller service so that each service's candidates are
analyzed together. This initial set is incomplete because an upstream loss can hide downstream breaks,
while changes to application and dependency code introduce new handoffs. Each
round therefore selects a bounded repair wave and re-queries the same metrics
after rollout. Verified repairs leave the set; remaining, newly exposed, and
newly introduced candidates form the next round.

\paragraphb{Reuse repairs without weakening validation.}
Engineers promote recurring validated repairs into knowledge-base rules:
recognition patterns, repair actions, and known-good wrappers or dependency
versions. A later case that matches a rule follows it instead of investigating
the same handoff again, which lowers cost and makes repairs of one kind more
uniform. Matching a rule does not show that the service is repaired. It must
still pass the validation in \S\ref{sec:design-validation}, and a failed case
returns to the next round.

\subsection{Why a Specialized Agentic System}
\label{sec:design-agent}

The specialized agentic system operationalizes the workflow above. The challenge lies
not in generating one context-propagation fix but in repeatedly combining program
understanding, controlled modification, runtime evidence, and coordination
across a changing fleet.

\paragraphb{Automate repeated program understanding.}
Engineers can repair individual cases and discover new patterns manually. The
bottleneck is recovering distributed knowledge for every case: starting from a
production-observed call, tracing unfamiliar application and dependency paths,
following explicit and shared-state handoffs, identifying their semantics,
preparing a narrow change, and explaining the evidence. Adding engineers does
not remove this per-case cost, while fixed scripts help only after a pattern
becomes a safe transform. \sys{} therefore automates the investigation and
repair before asking the owner to make decisions that require local authority.

\paragraphb{Specialize beyond a general coding agent.}
A general coding agent (e.g., Codex~\cite{openai2026codex} or
OpenCode~\cite{opencode2026}) is effective when given a
repository and a scoped task.
Broken propagation begins with a production symptom that has no source
location and ends only after deployment and re-measurement. It therefore
requires a specialized system whose workflow scopes the task before coding and
continues through runtime validation. \sys{} starts from a suspicious root-like
call, enumerates its execution paths,
locates the repair anchor, checks causality, lifetime, and attribution, and
either applies a supported repair or abstains. It then validates the change
with the same controlled traffic and retains the evidence across repositories
and repair rounds.

\paragraphb{Separate reasoning from control.}
The LLM agent supplies source code understanding and modification: it
follows unfamiliar paths, reasons about execution structure, applies a
supported local change, and explains it. The surrounding system
selects cases, constrains allowable repairs, preserves repair and validation
state, and connects each result to runtime evidence and production feedback.
The contribution is therefore not a better prompt or custom model, but a system
that turns an open-ended fleet problem into repair tasks that can be explained,
reviewed, measured, and repeated.

  \section{Implementation}
\label{sec:implementation}

\begin{figure}[t]
  \centering
  \resizebox{0.98\columnwidth}{!}{%
    % ============================================================
% Single-column implementation overview.
% palette: black + one accent (cfAccent); safe in greyscale
% needs: figures/context/preamble.tex
% ============================================================
\begin{tikzpicture}[
  font=\sffamily, text=black,
  box/.style={draw=black!55, fill=white, rounded corners=3pt, line width=0.5pt},
  core/.style={box, draw=black!80, line width=0.9pt},
  slot/.style={draw=black!55, densely dashed, rounded corners=2pt, fill=black!4,
               inner sep=1.5pt, font=\tiny, align=center},
  tuple/.style={draw=cfAccent, line width=0.9pt, fill=white, rounded corners=2.5pt,
                inner sep=2pt, font=\tiny, align=center},
  flow/.style={-{Stealth[length=1.5mm]}, line width=0.8pt, black!70},
  bidir/.style={{Stealth[length=1.5mm]}-{Stealth[length=1.5mm]}, line width=0.8pt, black!70},
  ret/.style={-{Stealth[length=1.5mm]}, line width=0.7pt, black!70, densely dashed},
  lab/.style={font=\tiny, text=black, inner sep=1pt, align=center},
  hd/.style={font=\tiny\bfseries, text=black, inner sep=1pt, align=center},
]

% ===================== input: the production tuple ====================
\node[tuple, minimum width=4.35cm] (tup) at (3.85,8.50)
  {\hspace{0.52cm}Target tuple: caller $\rightarrow$ callee $\cdot$ method};
\pic[scale=0.74] at ($(tup.west)+(0.25,0)$) {ico target};

% ===================== agent and supporting system ====================
\node[box, minimum width=2.35cm, minimum height=1.26cm] (skl) at (1.175,7.36) {};
\pic[scale=0.74] at (1.175,7.74) {ico book};
\node[hd]  at (1.175,7.45) {Skills \& Knowledge};
\node[lab] at (1.175,7.23) {Stage Order, Gates};
\node[lab] at (1.175,7.03) {Repair Taxonomy};

\node[core, minimum width=2.65cm, minimum height=1.26cm] (agt) at (3.875,7.36) {};
\pic[scale=0.78] at (3.875,7.67) {ico bot};
\node[hd] at (3.875,7.41) {Coding Agent};
\node[lab] at (3.875,7.19) {Harness \& Model};
\node[lab] at (3.875,6.99) {Interchangeable};

\node[box, minimum width=2.30cm, minimum height=1.26cm] (wsp) at (6.55,7.36) {};
\pic[scale=0.74] at (6.55,7.74) {ico server};
\node[hd]  at (6.55,7.45) {Durable Workspace};
\node[lab] at (6.55,7.23) {tasks.md, state/};
\node[lab] at (6.55,7.03) {repos/, report/};

\draw[flow]  (2.35,7.36) -- (2.55,7.36);
\draw[-{Stealth[length=1.1mm]}, line width=0.65pt, black!70]
  (5.18,7.43) -- (5.42,7.43);
\draw[-{Stealth[length=1.1mm]}, line width=0.65pt, black!70]
  (5.42,7.29) -- (5.18,7.29);
\draw[flow]  (tup) -- (agt);

% ===================== MCP tool plane =================================
\node[box, minimum width=7.70cm, minimum height=1.06cm] (mcp) at (3.85,5.84) {};
\pic[scale=0.70] at (0.28,6.16) {ico gear};
\node[hd, anchor=west] at (0.50,6.16) {MCP Server\;---\;System Access Layer};

\foreach \i/\ic/\tx in {0/glass/Discovery, 1/wrench/Build \& Deploy,
                        2/pulse/Replay \& Metrics, 3/gear/Deps \& Code}{
  \pic[scale=0.64] at ({0.95+1.90*\i},5.78) {ico \ic};
  \node[lab] at ({0.95+1.90*\i},5.52) {\tx};}

\draw[flow] (1.60,6.73) -- (1.60,6.37);
\node[lab, anchor=west] at (1.73,6.55) {Tool Calls};
\draw[flow] (6.10,6.37) -- (6.10,6.73);
\node[lab, anchor=east] at (5.97,6.55) {Structured Results};

% ===================== systems the tools act on =======================
\node[box, minimum width=3.40cm, minimum height=0.62cm] (env) at (1.70,4.66) {};
\pic[scale=0.78] at (0.32,4.66) {ico flask};
\node[lab, anchor=west] at (0.58,4.66) {Pre-production Testbed};

\node[box, minimum width=4.10cm, minimum height=0.62cm] (prd) at (5.65,4.66) {};
\pic[scale=0.78] at (3.92,4.66) {ico fleet};
\node[lab, anchor=west] at (4.18,4.66) {Production Fleet \& Telemetry};

\draw[flow] (1.60,5.31) -- (1.60,4.97);
\node[lab, anchor=west] at (1.73,5.14) {Deploy \& Replay};
\draw[flow] (6.10,4.97) -- (6.10,5.31);
\node[lab, anchor=east] at (5.97,5.14) {Idle vs.\ Load Metrics};

% ===================== campaign authority =============================
\node[box, minimum width=7.70cm, minimum height=0.84cm] (cmp) at (3.85,3.62) {};
\pic[scale=0.70] at (0.28,3.84) {ico scale};
\node[hd, anchor=west] at (0.50,3.84) {Campaign \& Engineering Authority};

\foreach \i/\ic/\tx in {0/funnel/Select Wave, 1/check/Owner Review,
                        2/wave/Roll Out, 3/eye/Re-measure}{
  \pic[scale=0.58] at ({0.45+1.90*\i},3.46) {ico \ic};
  \node[lab, anchor=west] at ({0.63+1.90*\i},3.46) {\tx};}

\draw[flow] (5.65,4.35) -- (5.65,4.04);
\node[lab, anchor=west] at (5.78,4.20) {Re-measure};

% ===================== the loop closes ================================
\draw[ret] (0.00,3.62) -- (-0.42,3.62) -- (-0.42,8.50) -- ($(tup.west)+(-0.03,0)$);
\node[lab, anchor=west] at (-0.38,8.78) {Next Wave};

% ===================== the system boundary ============================
\node[draw=black!35, densely dashed, rounded corners=5pt, line width=0.5pt,
      fit=(skl)(wsp)(mcp)(cmp), inner xsep=4pt, inner ysep=4pt] (per) {};

\end{tikzpicture}%
  }
  \caption{\sys{} implementation overview.}
  \Description{A coding agent is surrounded by repair-specific skills
  and knowledge, a durable workspace, and an MCP system-access layer connected to an
  isolated pre-production testbed and production telemetry. A campaign and
  engineering layer selects repair waves, reviews changes, rolls them out, and
  re-measures production.}
  \label{fig:implementation}
\end{figure}
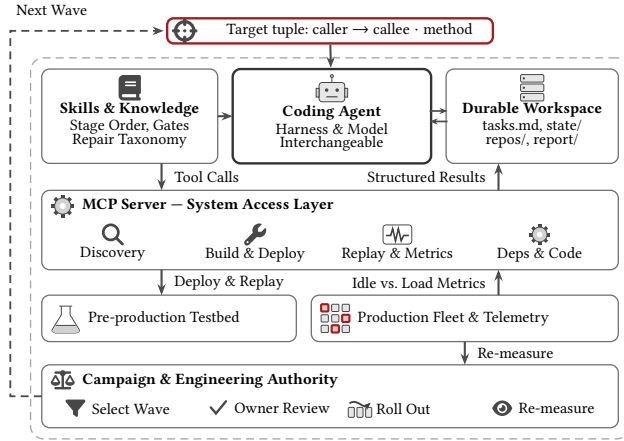

Figure~\ref{fig:implementation} shows our implementation around an
interchangeable coding agent. It has approximately 9,100 lines of Go code
for Model Context Protocol (MCP) integrations~\cite{anthropic2024mcp} and
32,400 lines of Skills and repair-knowledge
specifications.
Appendix~\ref{sec:appendix-implementation} provides additional implementation
details.

\subsection{Agent Architecture}
\label{sec:implementation-architecture}

\sys{} can use a tool-using coding agent, including Codex and OpenCode,
for understanding and modifying source code. The agent is not
given an unconstrained prompt for repair. Top-level Skills define the workflow stages
and exit conditions, while specialized Skills implement localization, repair, dependency updates, result audit, and
code-deploy validation. The Skills load a knowledge base in order of decreasing
generality, from general concepts through runtime and framework behavior to repair
playbooks, negative rules, and concrete version or wrapper indexes
(\S\ref{sec:implementation-repair-validation}). Because this knowledge supplies
the established propagation semantics and the known repairs, the model has to
explore only what is specific to the service.

Repair can span many execution paths, repositories, and runtime
experiments. To bound the main agent's context, \sys{} delegates
self-contained steps to subagents, including localization and
the build--deploy--replay--observe sequence. Each subagent receives only the
state needed for its step and returns a formatted result; the main agent
persists that result before advancing the workflow.

The MCP server connects the agent to operational systems through tools with
explicit input and output schemas. These tools collect runtime and service
metadata, inspect source code, update dependencies, and support build,
deployment, traffic replay, and code review. Together with the Skills,
this turns each workflow gate into a choice among admissible actions. Missing
or malformed fields trigger retry or an inconclusive result rather than an
arbitrary guess.
Appendix~\ref{sec:appendix-tools} lists the MCP tools exposed.

\sys{} keeps repair state outside the model context. Each service has a
workspace for its repositories, workflow progress, evidence, and reports,
while each observed call has a separate execution-path record. These records
connect a repair decision to the exact code and runtime experiment that support
it. The workflow checkpoints them after each consequential operation, so it can
pause and resume without repeating the investigation.

\subsection{Diagnose: Replay and Localize}
\label{sec:implementation-diagnosis}

Diagnosis starts from a production-observed call tuple: the \textless caller
service, callee service, callee method\textgreater{} defined in
\S\ref{sec:design-diagnose}. The telemetry tools retrieve the current root-like
calls, while the service-discovery tool resolves
the source repository, deployed version, and available pre-production clusters.
This preserves the production symptom as the anchor for both source analysis
and later validation.

For controlled traffic replay, \sys{} first deploys the pre-repair
code in an isolated test environment. It records a
baseline observation window with an idle workload and a replay window with the
target-request workload, then queries the same root-like call metrics.
Comparing the two windows determines whether the call changes with the target
request. The system preserves the workload for later validation.

Localization is deliberately read-only. For each propagation break, a localization subagent finds every accessible callsite for the
observed callee and method, then walks backward through application and dependency code.
It follows helpers, interfaces, callbacks, generated clients, goroutines,
channels, workers, and library boundaries, assisted by an asynchronous-node
scanner for common Go constructs. Each candidate path records its source
locations, repository owner, execution handoffs, how context is carried or
lost, and any unresolved source or dispatch edge. The result identifies the
last point with the required context and the first point where the work
proceeds without it, yielding the repair anchor. If the accessible source cannot establish
an anchor, the subagent records the missing evidence and abstains.

\subsection{Fix and Validate}
\label{sec:implementation-repair-validation}

\sys{} implements the handoff contracts in Figure~\ref{fig:repair-contracts}
through a knowledge base of recurring, validated repair patterns. Each pattern
connects source evidence at a repair anchor to the context that
must be restored, the component that owns the handoff, and a bounded change.
After localization, the agent selects a pattern whose conditions match the
complete execution path and applies it at the smallest boundary that both has the
required context and controls the receiving execution. This may change the
service, upgrade or patch a shared dependency, or adopt a wrapper. If
no pattern justifies the causality, lifetime, and attribution required by
\S\ref{sec:design-repair}, \sys{} abstains.
Representative patterns and exclusions appear in
Appendix~\ref{sec:appendix-knowledge}.

Loading the entire knowledge base into every session would crowd out the source
evidence, so the Skills load it in stages. The agent reads the small
framing documents once per service, condensing the stage order, the handoff
contracts, and the negative rules into a short checklist in the workspace that
cites the source section for each entry. The playbooks, version indexes, and
wrapper indexes are far larger, so the agent retrieves only the entries that the
current execution path matches. Each applied pattern records the rule it
followed for the audit to check.

We separate decision from modification. Read-only
localization records the repair anchor and its evidence; a write-capable stage
applies the selected pattern and records the diff. The workflow records each
claimed path, decision, and change in a fix log and structured workflow state.
A separate audit Skill treats these entries as claims: it reopens the cited source
locations and diff, then checks the stated cause and repair independently. The
audit therefore evaluates whether the recorded reasoning is supported, rather
than blindly accepting the repair summary or the final outcome as proof.

Validation reuses the baseline, replay workload, and observation parameters
recorded during diagnosis (\S\ref{sec:implementation-diagnosis}). The validation
Skill binds these inputs to the candidate commit and checkpoints the
resulting build, deployment, replay task, and metric window. This keeps
the code, deployment, workload, and measurement window fixed during the
comparison.

Validated repairs can extend the knowledge base. Engineers promote recurring
source patterns, exclusions, repair actions, fixed dependency versions, and
wrapper mappings for reuse by later services. Promotion remains gated by the
replay comparison, so a plausible source pattern is not generalized before its
effectiveness is observed at runtime.

\subsection{Cross-Repository and Fleet Operation}
\label{sec:implementation-fleet}

A repair may begin in an application repository and end in a shared executor,
SDK, or framework. The workspace carries the original call, path evidence,
selected handoff contract, and validation state across these repositories.
For an unreleased SDK patch, the system pushes a repair branch and pins the
consumer to an immutable candidate revision until an official release exists.
\sys{} builds both the original and modified dependency revisions to
verify that the patch does not break the build. The agent
preserves its source evidence, reasoning, diff, audit, and validation results in
the workspace, then uses MCP tools to create a merge request that repository
owners review, approve, and release manually.

At fleet scale, the campaign runs \sys{} separately for each service. Engineers
rank the current candidates and select a bounded wave; a batch runner then
creates an isolated workspace and executes the same evidence-gated workflow for
each service. Campaign state records whether each run is active, validated,
retryable, blocked, or awaiting review. Humans resolve permission, dependency,
review, and rollout decisions. After
owner-approved rollout, the campaign re-queries production telemetry and uses
the fixed, remaining, and newly visible calls to select the next wave.

\sys{} supports two supervision modes. In \emph{interactive mode}, the agent runs
in a coding-agent session, letting an engineer inspect evidence and
intervene at each gate. \emph{Batch mode} runs multiple isolated workflows
autonomously and pauses when evidence, permissions, review, or rollout requires human
judgment. Both modes use the same repair logic and validation standard; they
differ in supervision and throughput. \S\ref{sec:evaluation} reports the
repairs that both modes produced.

  \section{Evaluation}
\label{sec:evaluation}

We test whether \sys{} closes the local and fleet repair loops in
Figure~\ref{fig:design}. A successful repair must identify the intended
handoff contract and restore the missing causal chain without
hurting the downstream work. At fleet scale, the repair process must
remain effective across repository and release boundaries. We evaluate three
objectives.

\paragraphb{O1: End-to-end repair.} We measure whether deployed repairs restore
the missing causal chain while the downstream work continues to
execute, and whether \sys{} can produce the corresponding source-code repair
from the observed call tuple and the pre-repair source.

\paragraphb{O2: System specialization.} We separate the capability supplied by
\sys{} from that of the underlying coding agent and model, then measure how
much the localization, repair, and validation stages in
\S\ref{sec:design-one-break} each contribute.

\paragraphb{O3: Operational closure.} We evaluate whether repairs pass through
owner review, propagate across repository boundaries, remain effective after
rollout, and support repeated discovery of newly visible repair work.

\subsection{Methodology}
\label{sec:evaluation-method}

We use three production scopes, moving from broad production evidence to focused
analyses of repeated repair and individual rollouts.
\emph{(1) 27-scenario scope.} Before the evaluation, operators in one
region had selected 27 business-critical request scenarios for routine
monitoring of context propagation along downstream calls. Each scenario is a
large set of services reached from one entry service. Examples
include video playback, account login, social interactions, comment retrieval,
and personalized content feeds.
Appendix~\ref{sec:appendix-production-evaluation}
gives the complete anonymized list
and the microservice scale of each scenario. We freeze this list and
combine the downstream call graphs rooted at those entries, counting each
\textless caller service, callee service, callee method\textgreater{} tuple
once even when it belongs to several scenarios.
\emph{(2) Content-feed graph.} For one major scenario, we separately analyze
the downstream call graph for requests that serve or refresh the short video
platform's primary personalized content feed. This pre-existing operational
target, which we call the \emph{content-feed graph}, supports our study of
repeated repair.
\emph{(3) Rollout-aligned cohort.} We start from propagation breaks that were
observed at call tuples, addressed by repairs produced by \sys{},
and later verified as no longer present in production. We retain the
240 target tuples across 16 services for which we can also identify the
service's first production deployment after the repair was merged.

To evaluate O2, we build a controlled benchmark from 14 real services with
previously human-repaired propagation breaks. Each evaluated configuration
receives the call tuples at which those breaks were observed and the
pre-repair source, but no live deployment or production feedback. We use the human
repair as the reference and report \emph{targets reached}: the number of the
benchmark's 78 production-confirmed target tuples for which the resulting patch
edits a reference repair location. Reaching a target does not by itself make the edit correct. In the human references, these
targets correspond to 126 file changes spanning 32 repositories. We also
check whether the patch changes the same source location and leaves 8
additional tuples that require no changes untouched.
Appendix~\ref{sec:appendix-evaluation}
gives the benchmark construction, the four
unresolved call tuples excluded from these primary scores, and the complete
results.

\begin{figure}[t]
  \centering
  \resizebox{0.6\columnwidth}{!}{%
    % Deployment-aligned production results.
% Needs: figures/context/preamble.tex
\begin{tikzpicture}[
  font=\sffamily,
  axis/.style={black!65, line width=0.5pt},
  grid/.style={black!16, line width=0.35pt},
  present/.style={fill=black!58, draw=black, line width=0.5pt},
  broken/.style={
    fill=cfAccent!12,
    draw=black,
    line width=0.5pt,
    pattern={Lines[angle=45,distance=2.2pt,line width=0.45pt]},
    pattern color=black
  },
  lab/.style={font=\footnotesize, text=black, inner sep=0.5pt},
  value/.style={font=\footnotesize\bfseries, text=black, inner sep=0.5pt},
  boundary/.style={font=\scriptsize\bfseries, text=black, inner sep=1pt},
]

\def\ymax{1.80}
% Leave headroom above the 100% gridline for values on nearly full bars.
\draw[axis] (1.08,2.60) -- (1.08,0.52) -- (5.94,0.52);
\foreach \v in {0,25,50,75,100} {
  \pgfmathsetmacro{\y}{0.52+\v/100*\ymax}
  \draw[grid] (1.08,\y) -- (5.94,\y);
  \node[lab, anchor=east] at (1.00,\y) {\v\%};
}
\draw[black, densely dashed, line width=0.75pt]
  (2.68,0.48) -- (2.68,2.54);
\node[boundary, anchor=south] at (2.68,2.56) {deployment};
% Percentages read off the axis, so bar values carry no unit.
\foreach \x/\p/\b/\name in {
  1.88/98.57/90.46/{days $-14$--$-1$},
  3.48/98.51/8.70/{days 0--6},
  5.08/97.41/4.69/{days 7--30}} {
  \pgfmathsetmacro{\hp}{\p/100*\ymax}
  \pgfmathsetmacro{\hb}{\b/100*\ymax}
  \path[present] (\x-0.56,0.52) rectangle ++(0.32,\hp);
  \path[broken] (\x+0.24,0.52) rectangle ++(0.32,\hb);
  \node[value, anchor=south] at (\x-0.40,0.55+\hp) {\p};
  \node[value, anchor=south] at (\x+0.40,0.55+\hb) {\b};
  \node[lab, align=center] at (\x,0.25) {\name};
}
\path[present] (1.72,-0.13) rectangle ++(0.24,0.14);
\node[lab, anchor=west] at (2.04,-0.06) {call present};
\path[broken] (3.72,-0.13) rectangle ++(0.24,0.14);
\node[lab, anchor=west] at (4.04,-0.06) {propagation break};

\end{tikzpicture}%
  }
  \caption{Deployment-aligned production results.}
  \Description{The grouped bar chart shows call presence near 98 percent
  before and after deployment while the break rate falls from about 90 percent
  to below 5 percent.}
  \label{fig:evaluation-production}
\end{figure}
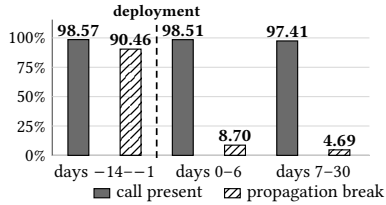

\subsection{Repairs Restore the Missing Causal Chain}
\label{sec:evaluation-local}

At the
latest complete snapshot on August 18, 2026, the 27-scenario scope
contains 26,136 observed call tuples from 2,587 services and carries
1.42 billion successful queries per second (QPS). The content-feed graph
accounts for 16,251 of these tuples, 1,528 services, and 1.13 billion QPS.

\paragraphb{Fleet-wide trend.} From April 28 to August 18,
2026, we measure the production signal over the 112 days with complete tuple
and traffic measurements. Following the propagation break definition in
\S\ref{sec:failure-semantics}, we report both the \emph{raw break rate} and the
\emph{actionable break rate}. The latter excludes tuples annotated as
known operational exceptions, such as periodic or sidecar-originated calls,
that should not trigger a repair.

To assess whether the trend spans different parts of the platform, we group
services by top-level product domain using the platform's internal
service-ownership metadata, and separately by primary implementation language. A
\emph{product domain} is a functional area such as content delivery, social
interaction, or account services.

In the 27-scenario scope, the raw tuple break rate falls from 32.52\% to 19.82\%,
while the actionable rate falls from 20.12\% to 8.55\%. Weighting tuples by
successful QPS gives corresponding declines from 19.68\% to 9.76\%
and from 6.90\% to 1.87\%, while observed services increase by 13.7\%. The content-feed
graph shows the same pattern: its raw and actionable tuple rates fall from
29.76\% to 19.51\% and from 16.38\% to 7.48\%, and its QPS-weighted rates fall
from 20.53\% to 9.99\% (raw) and from 5.61\% to 1.61\% (actionable), while
observed services increase by 6.1\%.
Among the 14 top-level product domains with at least 100 tuples at both
endpoints, 12 improve. All four groups defined by primary implementation
language also improve: Go, C++, Python, and Java.
The improvement therefore holds across a heterogeneous production graph that
keeps growing over the 112 days.

\paragraphb{Deployment-aligned effect.} The 240 target tuples correspond to
repairs produced by \sys{} and reviewed by repository owners. Production records
later report their breaks as no longer observed. For each
service, day 0 is the first successful production deployment recorded after
its repair was merged.
Figure~\ref{fig:evaluation-production} separates execution from observation:
telemetry contains 97.41--98.57\% of the expected daily observations
across the three measurement windows. Over the same windows,
the tuple break rate falls from 90.46\% to 8.70\% and then 4.69\%, and the
QPS-weighted break rate falls from 18.13\% to 0.60\% and then 0.42\%.
The transition is immediate at daily resolution, from 79.92\% on day $-1$ to
16.46\% on day 0 and 8.02\% on day 1, and all 16 services improve in both
post-deployment windows.

The preceding rates pool daily observations from all 240 target tuples, so a
service with more tuples contributes more. We therefore
also average the 16 service-level changes from the pre-deployment window equally.
The mean tuple break rate decreases by 75.12 percentage points during days
0--6 and by 80.50 percentage points during days 7--30. Resampling the 16
services gives 95\% confidence intervals of 61.69--87.16 and 70.04--89.94
percentage points, so the reduction remains large under different service mixes.
Both the pooled and equally weighted analyses show that the target calls
continue to appear in telemetry while their propagation breaks largely
disappear, matching the validation criteria in \S\ref{sec:design-validation}.
Appendix~\ref{sec:appendix-production-evaluation} gives the full longitudinal series,
the per-group breakdowns, and the measurement procedures.

\paragraphb{Benchmark evidence.} Each benchmark case supplies the
caller--callee--method tuple with the observed break and a frozen source
tree, then scores the resulting patch on two criteria: whether it repairs a
production-confirmed target and whether it changes the same source
location as the human repair. With Codex and GPT-5.4~\cite{openai2026gpt54},
the full system reaches all
78 production-confirmed targets and matches the human repair location in 77.
\S\ref{sec:evaluation-specialization} uses this benchmark to separate our system's
specialization from the coding agent and model.

\subsection{Specialization Drives Repair Capability}
\label{sec:evaluation-specialization}

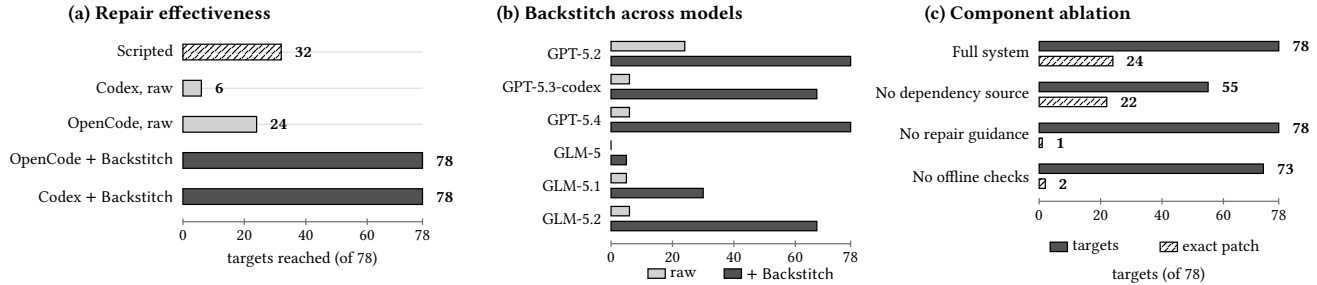
\begin{figure*}[t]
  \centering
  \resizebox{0.98\textwidth}{!}{%
    % Full-width controlled evaluation summary.
% Needs: figures/context/preamble.tex
\begin{tikzpicture}[
  font=\sffamily,
  axis/.style={black!55, line width=0.45pt},
  grid/.style={black!15, line width=0.35pt},
  raw/.style={fill=black!18, draw=black, line width=0.45pt},
  full/.style={fill=black!68, draw=black, line width=0.45pt},
  alt/.style={
    fill=white,
    draw=black,
    line width=0.45pt,
    pattern={Lines[angle=45,distance=1.8pt,line width=0.4pt]},
    pattern color=black
  },
  lab/.style={font=\tiny, text=black, inner sep=0.5pt},
  title/.style={font=\scriptsize\bfseries, text=black, anchor=west},
  value/.style={font=\tiny\bfseries, text=black, anchor=west},
]

\def\barmax{3.05}
\newcommand{\evalbar}[5]{%
  \pgfmathsetmacro{\w}{#4/78*\barmax}
  \node[lab, anchor=east] at (#1+1.48,#2) {#3};
  \draw[grid] (#1+1.58,#2) -- (#1+1.58+\barmax,#2);
  \path[#5] (#1+1.58,#2-0.10) rectangle ++(\w,0.20);
  \node[value] at (#1+1.63+\w,#2) {#4};
}

% Panel A: headline methods.
\begin{scope}
  \node[title] at (0,3.50) {(a) Repair effectiveness};
  \evalbar{0}{3.02}{Scripted}{32}{alt}
  \evalbar{0}{2.56}{Codex, raw}{6}{raw}
  \evalbar{0}{2.10}{OpenCode, raw}{24}{raw}
  \evalbar{0}{1.64}{OpenCode $+$ \sys{}}{78}{full}
  \evalbar{0}{1.18}{Codex $+$ \sys{}}{78}{full}
  \draw[axis] (1.58,0.86) -- (4.63,0.86);
  \foreach \v in {0,20,40,60,78} {
    \pgfmathsetmacro{\x}{1.58+\v/78*\barmax}
    \draw[axis] (\x,0.82) -- (\x,0.90);
    \node[lab, anchor=north] at (\x,0.78) {\v};
  }
  \node[lab] at (3.10,0.38) {targets reached (of 78)};
\end{scope}

% Panel B: matched models.
\begin{scope}[xshift=5.45cm]
  \node[title] at (0,3.50) {(b) \sys{} across models};
  \foreach \name/\rawv/\fullv/\yy in {
    GPT-5.2/24/78/3.00,
    GPT-5.3-codex/6/67/2.58,
    GPT-5.4/6/78/2.16,
    GLM-5/0/5/1.74,
    GLM-5.1/5/30/1.32,
    GLM-5.2/6/67/0.90} {
    \node[lab, anchor=east] at (1.48,\yy) {\name};
    \pgfmathsetmacro{\wr}{\rawv/78*\barmax}
    \pgfmathsetmacro{\wf}{\fullv/78*\barmax}
    \path[raw] (1.58,\yy+0.035) rectangle ++(\wr,0.12);
    \path[full] (1.58,\yy-0.155) rectangle ++(\wf,0.12);
  }
  \draw[axis] (1.58,0.58) -- (4.63,0.58);
  \foreach \v in {0,20,40,60,78} {
    \pgfmathsetmacro{\x}{1.58+\v/78*\barmax}
    \draw[axis] (\x,0.54) -- (\x,0.62);
    \node[lab, anchor=north] at (\x,0.50) {\v};
  }
  \path[raw] (2.05,0.14) rectangle ++(0.22,0.12);
  \node[lab, anchor=west] at (2.31,0.20) {raw};
  \path[full] (3.02,0.14) rectangle ++(0.22,0.12);
  \node[lab, anchor=west] at (3.28,0.20) {$+$ \sys{}};
\end{scope}

% Panel C: component ablation.
\begin{scope}[xshift=10.90cm]
  \node[title] at (0,3.50) {(c) Component ablation};
  \foreach \name/\covered/\exact/\yy in {
    Full system/78/24/3.00,
    No dependency source/55/22/2.48,
    No repair guidance/78/1/1.96,
    No offline checks/73/2/1.44} {
    \node[lab, anchor=east] at (1.48,\yy) {\name};
    \pgfmathsetmacro{\wc}{\covered/78*\barmax}
    \pgfmathsetmacro{\we}{\exact/78*\barmax}
    \path[full] (1.58,\yy+0.035) rectangle ++(\wc,0.12);
    \path[alt] (1.58,\yy-0.155) rectangle ++(\we,0.12);
    \node[value] at (1.63+\wc,\yy+0.095) {\covered};
    \node[value] at (1.63+\we,\yy-0.095) {\exact};
  }
  \draw[axis] (1.58,1.12) -- (4.63,1.12);
  \foreach \v in {0,20,40,60,78} {
    \pgfmathsetmacro{\x}{1.58+\v/78*\barmax}
    \draw[axis] (\x,1.08) -- (\x,1.16);
    \node[lab, anchor=north] at (\x,1.04) {\v};
  }
  \path[full] (1.72,0.48) rectangle ++(0.22,0.12);
  \node[lab, anchor=west] at (1.98,0.54) {targets};
  \path[alt] (3.12,0.48) rectangle ++(0.22,0.12);
  \node[lab, anchor=west] at (3.38,0.54) {exact patch};
  \node[lab] at (3.10,0.18) {targets (of 78)};
\end{scope}

\end{tikzpicture}%
  }
  \caption{Controlled repair results. (a) Representative execution methods.
  (b) Confirmed targets reached with and without \sys{} for six models.
  (c) Targets reached and exact matches to the human patch
  in the ablation study.}
  \Description{Three horizontal bar charts show repair effectiveness,
  raw-versus-specialized results for six models, and component
  ablations. In the ablation, the full system reaches all 78 confirmed targets
  and exactly matches the human patch at 24 targets. Removing dependency source trees
  reduces targets reached to 55, while removing repair guidance reduces exact
  matches to one.}
  \label{fig:evaluation-controlled}
\end{figure*}

The production measurements establish the scale and observed effect of
deployed repairs. The smaller controlled benchmark measures the
contribution of the design components in \S\ref{sec:design} to source-level
repair capability.

\paragraphb{Specialization across agents and models.}
Figure~\ref{fig:evaluation-controlled}(a) shows the representative GPT-5.4
conditions. Codex with the full \sys{} configuration
reaches all 78 confirmed targets and matches the human repair location in 77. The
same Codex--GPT-5.4 pair without \sys{} reaches 6. OpenCode improves from 24
targets to all 78 inside \sys{}, matching the human repair
location at all 78. As a non-agent comparison, a scripted fixer
encodes five common repair patterns from the same knowledge base as fixed
rewrites, checks each result with Go's parser and \texttt{gofmt}, and reaches
32. Every condition leaves all eight leave-unchanged cases untouched.
\sys{} supports interchangeable coding agents and models. To isolate
its contribution, each matched comparison holds the coding agent and model fixed and adds only
the system's repair specialization.
Figure~\ref{fig:evaluation-controlled}(b) reports both configurations per model:
GPT-5.2 rises from 24 to 78 targets, GPT-5.3-codex from 6 to 67,
GPT-5.4 from 6 to 78, GLM-5~\cite{zai2026glm5} from 0 to 5, GLM-5.1 from 5 to 30, and GLM-5.2
from 6 to 67. The specialization improves every matched pair.

\paragraphb{Component ablation.} Holding Codex and GPT-5.4 fixed,
Figure~\ref{fig:evaluation-controlled}(c)
removes one part of the local repair loop in \S\ref{sec:design-one-break} at a
time. First, localization must follow an
execution path across application and dependency code
(\S\ref{sec:implementation-diagnosis}). Removing access to dependency source
reduces targets reached from 78 to 55. This is the
largest loss because the observed call tuple does not reveal whether the
violated handoff belongs to application code, a dependency, or a shared
executor. Second, the repair stage uses knowledge-base guidance to connect a
localized handoff to an established repair
(\S\ref{sec:implementation-repair-validation}). Removing this guidance
does not affect localization: the agent still reaches all 78 targets because it
finds the reference locations. However, auditing each patch against the human
repair and the handoff contract finds clearly wrong repairs for at least 39 of
the 78 targets. Exact matches to the human patch also fall from
24/78 to 1/78. This metric measures how closely a patch follows an
already validated repair, not whether it is correct, because a handoff can be
repaired correctly in more than one way; the audit, not the
exact-match drop, shows these patches to be wrong.
Third, the benchmark can validate a repair without a live deployment only
through a candidate-commit build and source-level code checks. Removing these
checks reduces targets reached to 73.
Because the benchmark is offline, no condition receives
deploy--replay--observe feedback, so every result above measures raw repair
capability.
The complete ablation and scoring rules appear in
Appendix~\ref{sec:appendix-benchmark}.

\subsection{The Fleet Loop Sustains Repair}
\label{sec:evaluation-fleet}

\begin{figure}[t]
  \centering
  \begin{subfigure}[b]{0.49\columnwidth}
    \centering
    \resizebox{\linewidth}{!}{%
      % Repair outcomes recorded in each supervision mode.
% Needs: figures/context/preamble.tex
\begin{tikzpicture}[
  font=\sffamily,
  axis/.style={black!65, line width=0.5pt},
  grid/.style={black!16, line width=0.35pt},
  inter/.style={fill=black!58, draw=black, line width=0.5pt},
  batchbar/.style={
    fill=cfAccent!12,
    draw=black,
    line width=0.5pt,
    pattern={Lines[angle=45,distance=2.2pt,line width=0.45pt]},
    pattern color=black
  },
  lab/.style={font=\footnotesize, text=black, inner sep=0.5pt},
  value/.style={font=\footnotesize\bfseries, text=black, inner sep=0.5pt},
]

% Match the bounding box of the companion panel so both scale alike.
\path[draw=none] (0.19,-0.06) rectangle (5.08,3.24);

\def\xmax{2.90}
\def\vmax{3100}
\draw[axis] (1.55,3.24) -- (1.55,0.62) -- (4.55,0.62);
\foreach \v/\t in {1000/1k, 2000/2k, 3000/3k} {
  \pgfmathsetmacro{\x}{1.55+\v/\vmax*\xmax}
  \draw[grid] (\x,0.62) -- (\x,3.24);
  \node[lab, anchor=north] at (\x,0.54) {\t};
}
\node[lab, anchor=north] at (1.55,0.54) {0};

\foreach \yc/\v/\t in {2.72/1667/{1,667}, 1.86/442/442, 1.00/424/{424 (95.9\%)}} {
  \pgfmathsetmacro{\len}{\v/\vmax*\xmax}
  \path[inter] (1.55,\yc+0.03) rectangle ++(\len,0.28);
  \node[value, anchor=west] at (1.55+\len+0.06,\yc+0.17) {\t};
}
\foreach \yc/\v/\t in {2.72/3057/{3,057}, 1.86/280/280, 1.00/192/{192 (68.6\%)}} {
  \pgfmathsetmacro{\len}{\v/\vmax*\xmax}
  \path[batchbar] (1.55,\yc-0.31) rectangle ++(\len,0.28);
  \node[value, anchor=west] at (1.55+\len+0.06,\yc-0.17) {\t};
}

\node[lab, anchor=east] at (1.47,2.72) {tracked};
\node[lab, anchor=east] at (1.47,1.86) {labeled};
\node[lab, anchor=east] at (1.47,1.00) {repaired};

\node[lab, anchor=north] at (3.05,0.30) {breaks};

\path[inter] (3.10,1.80) rectangle ++(0.22,0.13);
\node[lab, anchor=west] at (3.38,1.865) {interactive};
\path[batchbar] (3.10,1.45) rectangle ++(0.22,0.13);
\node[lab, anchor=west] at (3.38,1.515) {batch};

\end{tikzpicture}%
    }
    \caption{Outcomes by mode.}
    \label{fig:evaluation-modes}
  \end{subfigure}
  \hfill
  \begin{subfigure}[b]{0.49\columnwidth}
    \centering
    \resizebox{\linewidth}{!}{%
      % Newly visible services over repeated repair rounds.
% Needs: figures/context/preamble.tex
\begin{tikzpicture}[
  font=\sffamily,
  axis/.style={black!65, line width=0.5pt},
  grid/.style={black!16, line width=0.35pt},
  frontier/.style={fill=black!42, draw=black!85, line width=0.4pt},
  trend/.style={cfAccent!88!black, line width=1.0pt},
  lab/.style={font=\footnotesize, text=black, inner sep=0.5pt},
  value/.style={font=\footnotesize\bfseries, text=black, inner sep=0.5pt},
]

\def\ymax{2.72}
\draw[axis] (0.94,0.52) -- (0.94,3.24) -- (5.08,3.24);
\foreach \v in {0,200,400,600} {
  \pgfmathsetmacro{\y}{0.52+\v/700*\ymax}
  \draw[grid] (0.94,\y) -- (5.08,\y);
  \node[lab, anchor=east] at (0.86,\y) {\v};
}
\foreach \round/\count/\x in {1/586/1.36,2/637/2.04,3/414/2.72,4/157/3.40,5/51/4.08,6/37/4.76} {
  \pgfmathsetmacro{\h}{\count/700*\ymax}
  \path[frontier] (\x-0.22,0.52) rectangle ++(0.44,\h);
  \node[value, anchor=south] at (\x,0.55+\h) {\count};
  \node[lab] at (\x,0.30) {\round};
}
\draw[trend, -{Stealth[length=1.5mm]}]
  (2.04,3.08) .. controls (2.74,2.75) and (3.42,1.18) .. (4.76,0.72);
\node[lab] at (3.02,-0.06) {repair round};
\node[lab, rotate=90] at (0.19,1.88) {new candidate services};

\end{tikzpicture}%
    }
    \caption{First-seen repair workload.}
    \label{fig:evaluation-frontier}
  \end{subfigure}
  \caption{The fleet loop.}
  \Description{Two bar charts. The left chart compares interactive and batch
  supervision on tracked, labeled, and repaired counts, with 95.9 percent and
  68.6 percent repaired shares. The right chart shows 586, 637, 414, 157, 51,
  and 37 newly visible services over six repair rounds.}
  \label{fig:evaluation-fleet}
\end{figure}
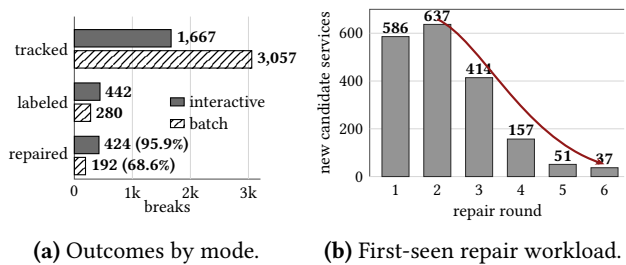

\paragraphb{Supervision modes.} Operators ran \sys{} in both modes of
\S\ref{sec:implementation-fleet}, and each mode maintained its own record of the
breaks it investigated (Figure~\ref{fig:evaluation-modes}). In interactive mode,
an engineer refines the diagnosis and inspects
the patch over several turns. This yields a high repaired share but
limits how many breaks the mode can process. Batch mode runs the same workflow
without engineer supervision, processing almost twice as many breaks but
repairing a smaller share of its deployed cases. The lower share shows the
remaining gap between engineer-guided repair and pure automation. Both modes
leave permission, review, and rollout to engineers.

\paragraphb{Owner review.} Owner review measures operational acceptance,
separately from source audit and runtime closure. Of
144 merge requests, owners reviewed 114 and approved 96 of them (84.2\%),
where approval means the request passed review or was merged.
Owners approve a repair to a shared library more often than
a repair to the caller service (94.1\% versus 76.2\%), and shared repairs
reach further: 66 of them address 123 calls across 33 caller groups and 47
repositories. Objections were rare: of 108 product-relevant
responses, 76 were positive or closed without objection and 25 asked for a
further check or an explanation. Only three (2.8\%) raised an explicit
negative or risk concern.\footnote{The remaining four responses were other or
unclear.} The audit (\S\ref{sec:implementation-repair-validation}) agreed with
production on 21 of 23; both misses were conservative rejections, so no wrong
fix reached a merge request.
Appendix~\ref{sec:appendix-adoption} gives the full
review, audit, and closure records and the cases outside current repair
scope; Appendix~\ref{sec:appendix-rollback} gives a case
study of a rollback that followed such a concern.

\paragraphb{Discovery-frontier contraction.} Repeated rounds test whether the
newly visible repair work from
\S\ref{sec:production-measurement} eventually contracts. Each round
selects services with a newly observed propagation break
that meets the campaign's traffic threshold, excluding those selected earlier.
The first two rounds in the content-feed graph find 586 and 637
such services as restored upstream context makes deeper calls observable and
topology changes introduce new work. The next four rounds find 414, 157, 51,
and 37. Figure~\ref{fig:evaluation-frontier} shows a 94.19\% contraction from
the round-2 peak. Across all rounds, the process identifies 1,882 distinct
candidate services, but rounds 5 and 6 contain dozens rather than
hundreds, turning first-seen discovery into routine follow-up.
Appendix~\ref{sec:appendix-frontier} gives the
selection rule and cumulative counts.

\paragraphb{Durability after rollout.} We exclude a 7-day rollout grace
period and count a recurrence only after telemetry reports the propagation
break on two consecutive days;
we omit calls that are no longer observable rather than counting their absence as success. Fifteen root-like calls
reappear, all from periodic, self-initiated, or uncontrolled refresh work.
Among calls whose repair should preserve request causality, none recur: 0/82
at 30 days and 0/78 at 60 days.
Appendix~\ref{sec:appendix-durability} gives the full
recurrence and observability rules.

\paragraphb{Operational cost.} Across 16 caller services
repaired in one recent campaign, the agent worked 9.99 hours in total and 21.95
minutes per service at the median.
In contrast, manual repair of a single break took 90 minutes at the median
(\S\ref{sec:production-measurement}). The campaign
consumed 169.75 million tokens, 90.78\% of which the model read from its
cache.
Appendix~\ref{sec:appendix-efficiency} gives the per-case timing and
token records.

Together, these results close the two loops in Figure~\ref{fig:design}. The
specialized system improves repair coverage across coding agents and models,
deployed repairs restore the missing causal record while preserving
the execution, and the fleet loop carries repairs through owner review and
adoption while repeated remeasurement reveals the next set of propagation
breaks.

  \section{Related Work}
\label{sec:related}

\paragraphb{Context propagation.} We compare \sys{} with the mechanisms that
prevent broken propagation in \S\ref{sec:existing-limits}: tracing
infrastructure~\cite{xtrace2007,sigelman2010dapper,sambasivan2016principled,
kaldor2017canopy}, transport
standards~\cite{w3ctracecontext,w3cbaggage,otelmessaging}, context
libraries~\cite{mace2018universal,opentelemetrycontext,
    opentelemetrypropagators}, mechanisms that also carry resource-control and
consistency state~\cite{mace2015retro,loff2023antipode}, framework middleware for
asynchronous paths~\cite{ravindranath2012appinsight,lin2018gammaray}, source
transformation along a configured path~\cite{welc2021contextpropagation}, and
sidecars that observe requests without inferring
causation~\cite{istiotracingfaq}. Each presumes the handoff is already
recognized. \sys{} complements them from the opposite end, beginning at a
production symptom that names no source location and discovering the handoff
none of them covers.

\paragraphb{Trace reconstruction.} Repairing a propagation break requires more than recovering the trace record. The missing parent--child
edges can be inferred offline from telemetry: Casper repairs inconsistent call graphs~\cite{huye2024casper},
TraceWeaver infers request trees from spans and
timing~\cite{ashok2024traceweaver}, and related systems recover request structure
without changing how the application propagates
context~\cite{shen2023deepflow,deeptrace2025,chainscope2026,reynolds2006pip,
tak2009vpath,chow2014mystery,zhang2023hindsight,ates2019pythia}. This restores
causality only in the record. Lifetime and attribution are properties of
the live execution (\S\ref{sec:failure-semantics}): a deadline and cancellation
scope must reach the worker while it runs, as must the identity and labels that
capacity planning and placement act on. An offline parent arrives too late to
govern the execution and cannot settle attribution, since a batch has no unique parent. Our system
repairs the handoff itself, so all three hold in the execution.

\paragraphb{Causal diagnosis.} Here, the call exists but its parent--child edge
is missing, so diagnosis proceeds by comparison. Prior work explains why an
event never happened by tracing the rules that should
have produced it~\cite{wu2014negative}, localizes a fault from the difference
between a working and a failing run~\cite{chen2016differential}, and carries
that difference into a candidate code change tested against historical
data~\cite{wu2017automated}. \sys{} applies a similar philosophy to context
propagation (\S\ref{sec:design-diagnose}). These systems stop at the responsible
rule or input, whereas our system must reach a handoff in a repository someone owns. Diagnosis and performance-analysis
systems~\cite{zhang2017pensieve,sun2023relational,exchain2024,crisp2022,
ikram2022causal,chen2024rcacopilot,autoarts2023,sieve2017,
hoffmann2018snailtrail} depend on accurate causal structure; \sys{} repairs the
propagation that supplies it.

\paragraphb{Automated repair and validation.} Industrial pipelines combine triage,
generation, retesting, and review at
scale~\cite{marginean2019sapfix,bader2019getafix,meta2026agenticrepair};
production-driven repair screens patches against replayed
traffic~\cite{durieux2017itzal,perkins2009clearview,arora2018replay,
antgroup2022replay}; and runtime evidence already constrains
repair~\cite{tracerepair2026,e2ereme2026}. Dr.Fix is closest in shape, repairing
Go concurrency defects with program analysis, tests, and owner
review~\cite{drfix2025}. ORCA repairs microservice incidents by diffing failure
telemetry against a healthy reference~\cite{orca2026}. Both start from a failure
oracle unavailable here: a race report names the defect and its evidence, and an incident supplies a
failing run to diff. A root-like call gives neither:
the request succeeds and all functional tests pass. A patch judged by so weak a signal can pass without
fixing anything~\cite{smith2015cure,rustassure2025}, so acceptance rests on
runtime validation of the deployed revision (\S\ref{sec:design-validation}).

\paragraphb{Coding agents and fleet change.} Our contribution is the system
surrounding the agent (\S\ref{sec:design-agent}). It specializes mechanisms
from three established lines of work. Repository agents and repair pipelines supply the
mechanics~\cite{xia2025agentless,
chen2024coder,huang2023agentcoder,yang2024sweagent,jimenez2024swebench,
zhang2024autocoderover,bouzenia2025repairagent,wang2025openhands}. Fleet-scale
generated change under staged review supplies the deployment
model~\cite{googlemigrations2025}. Governed execution supplies the gates:
reusable guidance needs applicability conditions and long trajectories need
audit~\cite{agentskills2026,agentrx2026}. STRATUS pairs specialized agents with
a transactional no-regression guarantee~\cite{stratus2025}, but mitigates a
running system rather than producing a reviewed source change remeasured
after rollout. Specialization, not
the coding agent or model, supplies our system's capability
(\S\ref{sec:evaluation-specialization}).

  \section{Conclusion}
\label{sec:conclusion}

A propagation break damages the record of an execution without stopping the
execution. \sys{} turns that surviving execution into the reference for repair,
from a production symptom through a bounded source change to a validating
replay. In production, the fleet's break rate more than halved over 112 days.
The same approach applies wherever a record breaks
while the behavior it describes survives. Frameworks prevent breaks on
recognized paths but cannot cover every handoff; for the paths they miss, a
specialized agentic system equipped with expert knowledge and runtime
validation makes repairing each site affordable.

  \bibliography{refs}

    \appendix
    \section{Implementation Artifacts}
\label{sec:appendix-implementation}

The implementation appendix presents selected artifacts from the system described in
\S\mainref{sec:implementation}. The excerpts are shortened and anonymized, but
preserve the field names, decisions, and gates used by the current system. We
include them to make the separation among agent reasoning, workflow policy,
repair knowledge, and runtime evidence concrete.

\subsection{Repository and Workspace Organization}
\label{sec:appendix-layout}

Figure~\ref{fig:repository-layout} shows the organization described in
the paper. The repository separates workflow policy (Skills), reusable repair
knowledge, and typed platform integrations. A service run creates a separate
workspace so that source checkouts and durable evidence are not stored in the
agent's conversation.

\begin{figure}[t]
\begin{minipage}{\columnwidth}
\scriptsize
\begin{Verbatim}[commandchars=\\\{\}]
\sys{}/
  skills/
    \sys{}/             # end-to-end service workflow
    service_info_collector/ # baseline and replay
    ctx-noise-handler/  # self-initiated calls
    ctx-async-handler/  # asynchronous boundaries
    ctxbreak-fix/       # per-call repair workflow
    ctxbreak-audit/     # read-only cause/fix audit
    result-audit/       # runtime closure audit
    code-deploy-verify/ # build, deploy, replay
    sdk-update-handler/ # dependency adoption
  knowledge/ctx_break_kb/
    L0-overview/        # routing and quick start
    L1-concepts/        # terms and taxonomy
    L2-principles/      # runtime/framework semantics
    L3-playbooks/       # repairs and anti-patterns
    L4-agent-workflow/  # evidence and stage protocol
    L5-reference/       # versions, wrappers, indexes
  mcp/tools/            # typed platform adapters

<service>.workspace/
  tasks.md              # stage and gate progress
  repos/                # application and dependencies
  state/                # resumable workflow records
  tmp/                  # checkpoints, scans, and logs
  report/               # audit and validation reports

<main-repo>/tmp/ctxbreak/
  state.json            # per-call workflow progress
  tuples/*.json         # per-call execution-path records
\end{Verbatim}
\end{minipage}
\caption{Abridged repository and per-service workspace layouts. Comments
summarize the responsibility of each directory.}
\Description{A directory tree separates agent Skills, a six-layer repair
knowledge base, and MCP tools. Service and repository workspaces store task,
checkpoint, execution-path, and report artifacts outside model context.}
\label{fig:repository-layout}
\end{figure}

The knowledge tree supports the handoff-contract decision in
\S\mainref{sec:design-repair}. After replay and source analysis identify a candidate
repair anchor, the agent loads the relevant runtime semantics, repair playbook,
and version or wrapper entries to determine the required causality, lifetime,
and attribution and select the smallest supported change. The workspace instead
persists the evidence produced across reproduction, localization, repair, and
validation, including checkpoints, execution paths, revisions, deployments, and
metric comparisons. This state remains available when the model context or
agent process ends, allowing the per-service loop to resume without repeating
completed steps or reconstructing evidence from conversation history.

\subsection{MCP Server}
\label{sec:appendix-tools}

Table~\ref{tab:mcp-tools} lists the external operations registered by the MCP
server. The interface keeps platform authentication and API details outside the
agent prompt. The tools expose what the agent can do, while the Skills specify
when the agent is allowed to use each operation.

\begin{table*}[t]
  \centering
  \begin{tabular}{@{}p{0.19\textwidth}p{0.45\textwidth}p{0.29\textwidth}@{}}
    \toprule
    \textbf{Capability} & \textbf{Representative registered tools} &
    \textbf{Recorded result} \\
    \midrule
    Telemetry &
      \path{get_production_broken_calls},
      \path{get_test_broken_calls} &
      Production candidates and control/experiment call sets. \\
    Service discovery &
      \path{get_service_metadata} &
      Repository, deployed version, clusters, and resource configuration. \\
    Source and repair &
      \path{find_asynchronous_handoffs}, \path{repair_handoff},
      \path{repair_call_chain} &
      Candidate asynchronous nodes, proposed changes, and dry-run output. \\
    Dependencies &
      \path{get_dependency_upgrade}, \path{update_dependencies} &
      Selected module versions and applied dependency updates. \\
    Build and deployment &
      \path{create_build}, \path{get_build_status},
      \path{get_build_log}, \path{deploy_service},
      \path{get_deployment_status} &
      Exact build version, failure log, deployment identifier, and status. \\
    Traffic replay &
      \path{create_replay_case}, \path{start_traffic_replay},
      \path{get_replay_status} &
      Replay case, task identifiers, and completion status. \\
    Repository handoff &
      \path{request_repository_access},
      \path{create_or_get_merge_request} &
      Access result and the created or reused review artifact. \\
    \bottomrule
  \end{tabular}
  \caption{Representative tools registered by the MCP server in the current
  implementation.}
  \label{tab:mcp-tools}
\end{table*}

\subsection{Skill-Governed Agent Execution}
\label{sec:appendix-skill}

\sys{} implements an end-to-end service repair workflow over the tools,
knowledge, and evidence gates described above. It maintains a dedicated
workspace for service-level progress, analyzes each observed call separately,
and validates the accumulated changes after each phase. The workflow does not
prescribe a fixed sequence of queries. Instead, its Skills constrain
when the coding agent may move from investigation to modification.
Table~\ref{tab:skill-gates} condenses this protocol. Localization and audit are
read-only; only the repair stage can modify a repository. This prevents a
plausible patch from becoming its own evidence that the diagnosis was correct.

The main agent also delegates self-contained steps to subagents so that a long
repair does not require one model context to retain every search and runtime
operation. A localization subagent handles one observed call, while validation
subagents perform bounded deployment, traffic replay, and observation steps.
Each receives only its task-local inputs and must return one schema-valid
record. The main agent serializes these records into the workspace before it
checks the next gate; subagents do not share their raw context with one another.

Subagent outputs are deliberately narrow. Localization returns one of
\texttt{Matched}, \texttt{Not\_Matched}, or \texttt{Inconclusive}, together
with an admitted repair action when the evidence supports one. Validation
returns bounded status values and the identifiers needed to reproduce each
operation. The main agent checks these fields against the current gate and is
the only writer of shared workflow state. If an output is malformed or omits a
required field, the workflow retries the task once and then records it as
\texttt{Inconclusive}. This interface turns open-ended exploration inside a
subagent into a bounded decision at the workflow boundary.

\begin{table}[t]
  \centering
  \begin{tabular}{@{}p{0.17\columnwidth}p{0.25\columnwidth}p{0.48\columnwidth}@{}}
    \toprule
    \textbf{Stage} & \textbf{Authority} & \textbf{Required artifact or gate} \\
    \midrule
    Localize & Read only & Candidate execution paths, repair anchors, and
      explicit unresolved edges. \\
    Choose contract & No source writes & Workflow checks every path against the
      repair families, recognition rules, and exclusions. \\
    Repair & Scoped writes & One admitted repair action and the resulting diff,
      tied to its repository and revision. \\
    Audit & Read only & Separate verdicts for the repair anchor and handoff
      contract, and for whether the diff restores that contract. \\
    Validate & Operational tools & Exact build and deployment provenance, replay
      identifiers, metric windows, and the comparison with the baseline. \\
    \bottomrule
  \end{tabular}
  \caption{Abridged execution gates from the repair Skills.}
  \label{tab:skill-gates}
\end{table}

The two audit Skills close different gates. For each path, the fix log and
structured workflow record preserve the observed call, source chain, claimed
repair anchor, supporting source locations, selected handoff contract, repair
target and action, and any evidence gaps. The source-and-diff audit treats
these fields as the repair agent's claims. It reopens the cited source and diff
to check the claimed cause and repair independently; a runtime result does not
by itself make the recorded reasoning valid.
\path{result-audit} repeats the runtime comparison and classifies propagation
breaks as unresolved or regressed before declaring full, partial, or failed closure.
The end-to-end workflow updates \texttt{tasks.md} after every task and resumes from
the first incomplete entry; a failed phase gate blocks later phases. Validation
reports retain a snapshot of this checklist with the branch and commit, which
prevents a cumulative result from being attributed to the wrong phase.

Figure~\ref{fig:skill-excerpt} shows an abridged excerpt from the read-only
localization prompt, named \path{subagent-attribution} in the repository. We
omit repository-specific commands and repeated field definitions. The excerpt
shows how missing source, unresolved dispatch, or an unsupported handoff contract
becomes an \texttt{Inconclusive} case while other calls continue independently.

\begin{figure}[t]
\begin{minipage}{\columnwidth}
\scriptsize
\begin{verbatim}
Localization subagent constraints:
1. Find the complete execution path before choosing
   a repair anchor. Record each path separately.
2. Start in the service repository. Follow helpers,
   interfaces, vendor code, and resolved modules.
3. Every hop must include repository, file, line,
   and function evidence.
4. Choose a contract only with a supported knowledge rule.
   Otherwise return Inconclusive and evidence_gap.
5. Do not modify code. Return exactly one structured
   result for the workflow.
\end{verbatim}
\end{minipage}
\caption{Abridged excerpt from the read-only localization prompt.}
\Description{The Skill requires complete execution-path evidence, dependency
inspection, a knowledge-supported handoff contract, explicit inconclusive results, and
prohibits source modification.}
\label{fig:skill-excerpt}
\end{figure}

\subsection{Recording Candidate Execution Paths}
\label{sec:appendix-attribution}

\S\mainref{sec:design-diagnose} describes how \sys{} walks backward from an
observed downstream call to the request or background activity that produced
it. The implementation records each candidate execution path in the
\texttt{chains} array: an ordered sequence of source locations that connects a recognizable
entry point, such as a handler or worker, to the target call across function,
queue, callback, and dependency boundaries. Each path must identify the last
location with the required context and the first representation of the work
without it. These locations establish the repair anchor; the intervening hops
show that the anchor can actually lead to the observed call. Because several
call sites or dispatch paths may produce the same call, \sys{} retains one
record for each supported path rather than selecting the first plausible
path.

Figure~\ref{fig:attribution-schema} shows the record that carries these paths
from diagnosis into repair. Each element of \texttt{chains} names the path's
entry, lists its ordered source locations, explains the repair anchor, and
identifies the component that owns the handoff. The record also cites the
knowledge rule used to interpret the path and notes any unresolved source or
dispatch edge. These fields allow the next gate to check both parts of the
design argument: that the path explains the runtime symptom and that its
causality, lifetime, and attribution justify a handoff contract from
\S\mainref{sec:design-repair}.

\begin{figure}[t]
\begin{minipage}{\columnwidth}
\scriptsize
\begin{verbatim}
{
  "observed_call": {
    "caller_service": "<caller>",
    "callee_service": "<callee>",
    "callee_method": "<method>"
  },
  "is_chain_found": true,
  "chains": [{
    "entry": "handler|worker|fan_in|...",
    "chain_detail": "repo:file:line -> ...",
    "classification":
      "Matched|Not_Matched|Inconclusive",
    "subclass": "R1.1|...|R4.1|null",
    "evidence": "source locations and reason",
    "kb_reference": "rule and exclusions",
    "fix_owner": "repo|sdk|both",
    "recommended_fix_path": "<repair action>"
  }],
  "call_decision": {
    "fix_path": "<repair action>",
    "fix_target": "module, version, or location",
    "rationale": "why this action applies"
  },
  "evidence_gap": []
}
\end{verbatim}
\end{minipage}
\caption{Abridged output schema from the read-only localization prompt.}
\Description{An abridged JSON object records the observed caller, callee, and
method; candidate execution paths and their status; the selected repair action;
and any unresolved evidence gaps.}
\label{fig:attribution-schema}
\end{figure}

The path record is a claim backed by inspectable source, not a summary that the
repair stage must trust. Each hop names a repository-relative file, line, and
function so the audit can reopen the same locations. The internal
\path{subclass} field selects a recognition rule that refines one repair
family from Figure~\mainref{fig:repair-contracts}; \path{classification} records
whether the path evidence satisfies that rule. A matched status means that the
source evidence and rule support the proposed handoff contract. A non-matching
status rejects that contract for the path, while an inconclusive status
preserves a path whose evidence is incomplete. After
checking all paths, \sys{} writes the selected action and target to
\path{call_decision}. If no path supports a repair, the evidence gaps remain
explicit and \sys{} abstains.

\sys{} stores each detailed path record immediately and maintains a smaller
service-level index for resumption. Figure~\ref{fig:service-state} shows this
index: it identifies the service and branch, records the current phase and
shared prerequisite changes, and links each observed call to its own evidence
record and status. Separating the index from the detailed evidence lets the
workflow resume at the first incomplete call without loading every execution path
into the model context. The \texttt{session\_lock} binds an agent session to one service. A resumed run
recovers that identity from the saved index rather than scanning other
workspaces and inferring which service to continue. This prevents concurrent
repairs from mixing execution-path evidence or repository changes.

\begin{figure}[t]
\begin{minipage}{\columnwidth}
\scriptsize
\begin{verbatim}
{
  "service": "<service>",
  "session_lock": "<agent-session>",
  "branch": "<repair-branch>",
  "phase": "source_localization",
  "calls": [{
    "idx": 1,
    "parent_record_id": "<record>",
    "callee_service": "<callee>",
    "callee_method": "<method>",
    "status": "localized"
  }],
  "baseline_actions": {
    "tracing_sdk_upgraded": false,
    "set_default_async_child": false,
    "sdk_batch_upgrade": "pending"
  }
}
\end{verbatim}
\end{minipage}
\caption{Abridged per-service durable state from the top-level repair Skill.}
\Description{A JSON record identifies the service, branch, workflow phase,
agent session, per-call progress, and prerequisite tracing actions.}
\label{fig:service-state}
\end{figure}

\subsection{Knowledge Rules and Negative Constraints}
\label{sec:appendix-knowledge}

The knowledge base records applicability conditions rather than a collection
of past patches. Table~\ref{tab:kb-examples} gives four representative rules
from the Go knowledge tree. These examples are useful because superficially
similar worker executions require different parents and lifetimes.

\begin{table*}[t]
  \centering
  \begin{tabular}{@{}p{0.20\textwidth}p{0.29\textwidth}p{0.27\textwidth}p{0.18\textwidth}@{}}
    \toprule
    \textbf{Observed execution} & \textbf{Required evidence} &
    \textbf{Admitted repair} & \textbf{Key exclusion} \\
    \midrule
    One request task crosses a channel & The submitted task and worker form a
      one-to-one request continuation, and the handoff fails to carry or restore
      the request state. & Wrap the task at submission and restore it only while
      consuming that task. & Do not apply when several requests contribute to
      one operation. \\
    Fan-in or delayed batch & Multiple request inputs contribute to one worker
      operation, so no single input is a truthful parent. & Create an explicit
      fan-in root rather than inheriting one request. & Do not select an
      arbitrary request context. \\
    Message-consumer entry & The messaging framework created the consumer span,
      but worker execution did not restore it into the execution-local Context.
      & Restore the framework context at the worker entry. & Do not relabel the
      consumer as periodic work or create a competing root. \\
    Finished-parent timing case & Context already crosses the handoff, but the
      tracing version and configuration detach a child created after its parent
      finishes. & Upgrade the tracing SDK and enable its asynchronous-child
      behavior. & Do not use this upgrade to conceal an unfixed application
      handoff. \\
    \bottomrule
  \end{tabular}
  \caption{Representative knowledge-base rules. Each rule combines positive
  evidence, a repair, and an exclusion.}
  \label{tab:kb-examples}
\end{table*}

Figure~\ref{fig:knowledge-excerpt} shows an abridged playbook excerpt that
combines a positive rule with an exclusion. The labels are stable identifiers
referenced by the execution-path records and audit outputs. Negative rules are
checked before modification and again during audit. Version indexes use the
same evidence-oriented form: each entry records an affected
version range, a confirmed fixed version, the issue type, and evidence for the
mapping. These constraints let later cases reuse a repair without reducing the
handoff-contract decision to syntax matching.

\begin{figure}[t]
\begin{minipage}{\columnwidth}
\scriptsize
\begin{verbatim}
T-3C  ch <- Task{Req: req}; Task has no context;
       a separate worker later issues the RPC.
       If the task is a 1:1 request continuation,
       use WrapRequestTask + HandleRequestTask.
       If it is fan-in, create a fan-in root.

W-11  Do not repair a request-scoped worker with
       DoPeriodicCall or DoFanInCall. Doing so hides
       the real request origin behind a custom root.
\end{verbatim}
\end{minipage}
\caption{Abridged excerpt from the playbook.}
\Description{A positive rule classifies a channel handoff according to whether
it is a one-to-one continuation or fan-in. A negative rule forbids replacing a
request's causal chain with a periodic or fan-in root.}
\label{fig:knowledge-excerpt}
\end{figure}

\subsection{Checkpointed Validation Evidence}
\label{sec:appendix-validation}

Runtime validation uses a step-indexed state file. A step becomes complete only
after its output is written. Figure~\ref{fig:validation-checkpoint} shows how
the retained state lets a run resume and binds the metric result to the
candidate and baseline environments under comparison. The full validation state also keeps a timestamped operation history.
The final report adds both call sets and classifies propagation breaks as
repaired, remaining, or newly observed. A later audit uses the latest comparison to
classify the overall result as full, partial, or failed closure. The retained
records expose both that verdict and the operations that produced it.

\begin{figure}[t]
\centering
\begin{minipage}{\columnwidth}
\scriptsize
\begin{verbatim}
{
  "steps": {
    "build_wait":     {"status":"completed"},
    "test_deploy":    {"status":"completed"},
    "metrics_compare":{"status":"pending"}
  },
  "context": {
    "candidate_environment": "<candidate-test>",
    "control_environment": "<baseline-test>",
    "source_commit": "<commit>",
    "build_id": "<build>",
    "deployment_id": "<deployment>",
    "replay_task_ids": ["<replay-task>"]
  }
}
\end{verbatim}
\end{minipage}
\caption{Abridged validation checkpoint. The environments and identifiers bind
the comparison to the exact code and workload under test.}
\Description{A JSON checkpoint records completed build and deployment steps,
the pending metric comparison, and identifiers for the candidate and baseline
environments, source commit, build, deployment, and traffic replay.}
\label{fig:validation-checkpoint}
\end{figure}

    \section{Evaluation Details}
\label{sec:appendix-evaluation}

This appendix gives the benchmark construction rules, complete comparison
tables, and supporting denominators for \S\mainref{sec:evaluation}. Each subsection
identifies the records it includes and the unit it counts. The production
measurements, rollout-aligned analysis, governance records, repeated
measurements of the content-feed graph, and offline benchmark are separate
datasets, not successive subsets of a single dataset.

\subsection{Controlled Benchmark}
\label{sec:appendix-benchmark}

The benchmark starts from a fixed pool of propagation-break cases that engineers had
already investigated. A case is included when it has a human repair and a
reference diff. At each of its primary target tuples, production measurement
shows that the propagation break is no longer observed, and the human change
repairs the break. The
resulting benchmark represents 14 real services.
For each service, it contains sanitized caller--callee--method tuples and frozen
source trees containing only repository history available before the reference
repair. The agent receives no live deployment or production feedback.

The manual references span 32 repository-level diffs and 126 files. Scoring
uses 90 call tuples: 78 at which production measurement confirmed the human
repair, four at which a propagation break remained and a source repair was
still possible, and
eight for which the benchmark requires no source change. The primary metric
counts how many of the 78 confirmed targets a patch reaches at a
reference repair location; it does not by itself establish that the edit is
correct. The
repair-location metric counts a patch as a match when it changes the same
source location as the reference human repair, even when its diff is not
textually identical. Exact diff equality is reported
only as a patch-shape diagnostic. A separate column counts how many of the four
tuples that remained broken a patch also edits; the human
reference edits three of the four. Several call tuples can share one cause,
patch, merge request, or deployment.

The scripted fixer is a deterministic, non-agent implementation of five
repair patterns from the knowledge base. Its Go rewrites pass an in-scope
request context through calls, replace visible-method uses of
\texttt{context.Background}, repair selected goroutine handoffs, replace known
wrapper imports, and enable asynchronous child spans when required. It parses
and formats every modified file and reverts a rewrite if either check fails.

For the full system and the three ablations in
Table~\ref{tab:appendix-ablation}, the exact-diff counts are, respectively,
24/78, 22/78, 1/78, and 2/78. The low value without repair knowledge
shows that reaching a reference repair location does not imply a correct edit;
the static audit below separates demonstrably wrong edits from allowed
alternatives. All completed benchmark conditions
leave the eight leave-unchanged call tuples alone. Adding \sys{} raises the
number of confirmed targets a model reaches by 54 for GPT-5.2, 61 for
GPT-5.3-codex, 72 for GPT-5.4, five for GLM-5, 25 for GLM-5.1, and 61 for
GLM-5.2.

\begin{table*}[t]
  \centering
  \begin{tabular}{@{}lllrrrr@{}}
    \toprule
    \textbf{Agent} & \textbf{Model} & \textbf{\sys{} support} &
      \textbf{Targets} & \textbf{Unresolved} & \textbf{Exact match} &
      \textbf{Location match} \\
    \midrule
    Scripted fixer & None & No & 32/78 & 0/4 & 7/78 & 32/78 \\
    Codex & GPT-5.4 & No & 6/78 & 0/4 & 0/78 & 6/78 \\
    OpenCode & GPT-5.4 & No & 24/78 & 0/4 & 3/78 & 24/78 \\
    OpenCode & GPT-5.4 & Full & \textbf{78/78} & 2/4 & 34/78 &
      \textbf{78/78} \\
    Codex & GPT-5.4 & Full & \textbf{78/78} & 1/4 & 24/78 &
      77/78 \\
    \bottomrule
  \end{tabular}
  \caption{Completed headline benchmark conditions. The unresolved column
  counts how many of the four tuples that remained broken a patch edits.}
  \label{tab:appendix-benchmark-headline}
\end{table*}

\begin{table*}[t]
  \centering
  \begin{tabular}{@{}lrrr@{\hspace{18pt}}rrr@{}}
    \toprule
    & \multicolumn{3}{c}{\textbf{Raw agent}} &
      \multicolumn{3}{c}{\textbf{With \sys{}}} \\
    \cmidrule(lr){2-4}\cmidrule(l){5-7}
    \textbf{Model} & \textbf{Targets} & \textbf{Exact match} &
      \textbf{Location match} & \textbf{Targets} & \textbf{Exact match} &
      \textbf{Location match} \\
    \midrule
    GPT-5.2 & 24/78 & 0/78 & 24/78 & 78/78 & 22/78 & 78/78 \\
    GPT-5.3-codex & 6/78 & 0/78 & 6/78 & 67/78 & 5/78 & 67/78 \\
    GPT-5.4 & 6/78 & 0/78 & 6/78 & 78/78 & 24/78 & 77/78 \\
    GLM-5 & 0/78 & 0/78 & 0/78 & 5/78 & 0/78 & 5/78 \\
    GLM-5.1 & 5/78 & 0/78 & 5/78 & 30/78 & 0/78 & 30/78 \\
    GLM-5.2 & 6/78 & 0/78 & 6/78 & 67/78 & 1/78 & 67/78 \\
    \bottomrule
  \end{tabular}
  \caption{Matched model comparisons. Figure~\mainref{fig:evaluation-controlled}(b)
  plots the target columns; the appendix also reports exact patch matches and
  matches to the human repair location.}
  \label{tab:appendix-backends}
\end{table*}

\begin{table}[t]
  \centering
  \begin{tabular}{@{}l@{\hspace{5pt}}r@{\hspace{5pt}}r@{\hspace{5pt}}r@{}}
    \toprule
    \textbf{Removed} & \textbf{Targets} & \shortstack{\textbf{Exact}\\\textbf{match}} &
      \shortstack{\textbf{Location}\\\textbf{match}} \\
    \midrule
    None & 78/78 & 24/78 & 77/78 \\
    No dependency source & 55/78 & 22/78 & 55/78 \\
    Repair guidance & 78/78 & 1/78 & 78/78 \\
    Offline checks & 73/78 & 2/78 & 73/78 \\
    \bottomrule
  \end{tabular}
  \caption{Component ablation with fixed Codex and GPT-5.4.}
  \label{tab:appendix-ablation}
\end{table}

The three ablations correspond to localization, repair, and validation in
\S\mainref{sec:design-one-break}. \emph{No dependency source} removes all
materialized dependency and asynchronous-library source trees, restricting
localization to each service's application source
(\S\mainref{sec:implementation-diagnosis}). \emph{Repair guidance} removes the
knowledge base and its analysis procedures from
\S\mainref{sec:implementation-repair-validation} while retaining the surrounding
workflow. \emph{Offline checks} removes the candidate-commit build and
source-level code checks from that validation procedure. Since the benchmark
is offline, it does not perform the deploy--replay--observe comparison. The
production experiment in \S\mainref{sec:evaluation-local} evaluates the deployed
replay comparison specified in
\S\mainref{sec:design-validation}.

Removing repair guidance changes what the patches contain, not where they land.
Targets reached and location match remain 78/78; all eight leave-unchanged
tuples remain untouched. The ablation therefore exposes a repair-selection
failure rather than missed localization.
We audit the edits against the frozen source, human repair, and
handoff contract. The audit counts an edit only when this static evidence proves
it wrong; it excludes omissions, different but allowed repairs, unanchored
edits, and cases that require runtime validation.
Among the 14 reference-comparable service tasks, 12 contain at least one clearly wrong
edit, for 26 distinct wrong repair decisions. Ten of the 78 primary targets map
to periodic loops repaired with \texttt{ctxgov.Go} instead of a per-iteration
\texttt{DoPeriodicCall}. Another 29 map to incomplete tracing-library repairs
that enable asynchronous child spans without raising the library to the
required version. These sets are disjoint, so the conservative count is 39/78
targets.

Because the benchmark supplies the call tuple at which the break was observed,
it does not test production
candidate discovery and only partially exercises diagnosis. It tests source
localization, repair synthesis, analysis across repositories, and the offline
validation procedure under fixed inputs.

\subsection{Production Measurement and Deployment Indexing}
\label{sec:appendix-production-evaluation}

The production scope is based on a fixed list of 27 business-critical request
scenarios from a pre-existing governance registry in one region. Each scenario
has an entry service and method. We join its
downstream service membership to daily call telemetry, then deduplicate on
caller service, callee service, and callee method so that a tuple reachable
from several scenarios is counted once. The analysis includes observed
services that execute application code and excludes storage, middleware,
sidecars, and other infrastructure targets.

At the latest complete snapshot, August 18, 2026, this scope contains 26,136
call tuples, 2,587 observed services, and 1.42 billion successful
queries per second (QPS). Table~\ref{tab:appendix-production-scenarios} gives
the frozen manifest in anonymized form and reports the number of distinct
downstream services observed for each scenario. Because a service can
appear downstream of several scenarios, these counts overlap and do not sum to
the 2,587-service union. The registered reply-listing scenario had no separately
labeled downstream membership record in this snapshot and therefore has a
count of zero. The content-feed graph is one scenario in this manifest; it
contains 16,251 tuples, 1,528 observed services, and 1.13 billion QPS.

\begin{table*}[t]
  \centering
  \setlength{\tabcolsep}{5pt}
  \begin{tabular}{@{}r l r@{\hspace{18pt}}r l r@{}}
    \toprule
    \textbf{ID} & \textbf{Request scenario} & \textbf{Services} &
    \textbf{ID} & \textbf{Request scenario} & \textbf{Services} \\
    \midrule
    01 & Location authorization & 115 &
    15 & Account cancellation & 277 \\
    02 & Video playback & 38 &
    16 & Inbox notification count & 69 \\
    03 & Personalized content feed & 1,528 &
    17 & User blocking & 140 \\
    04 & Account-region lookup & 25 &
    18 & User following & 321 \\
    05 & User login & 205 &
    19 & Following-list retrieval & 169 \\
    06 & Account switching & 182 &
    20 & Follower-list retrieval & 70 \\
    07 & Federated login & 398 &
    21 & Like or unlike & 220 \\
    08 & Email registration or login & 220 &
    22 & Profile post list & 64 \\
    09 & Quick login & 166 &
    23 & Other-user profile & 255 \\
    10 & SMS-code request & 86 &
    24 & Comment listing & 157 \\
    11 & QR-code login & 33 &
    25 & Reply listing & 0 \\
    12 & Identity-token login & 235 &
    26 & Repost consumption & 98 \\
    13 & Passkey registration & 80 &
    27 & Lightweight-app feed & 1,174 \\
    14 & Account logout & 125 & & & \\
    \bottomrule
  \end{tabular}
  \caption{Anonymized manifest of the 27 business-critical request scenarios.
  Service counts are distinct downstream services observed on August
  18, 2026. Counts overlap across scenarios.}
  \label{tab:appendix-production-scenarios}
\end{table*}

For each observed call tuple, telemetry records whether context continuity is
observed across the call. The raw tuple break rate divides the tuples without
observed continuity by all observed tuples. The actionable break rate excludes
tuples with a named operational exception, including periodic, self-initiated,
sidecar-originated, and other known non-actionable patterns. The QPS-weighted
rates divide the successful QPS carried by tuples without observed continuity
by the total successful QPS.

April 28 is the earliest date with complete and comparable tuple and QPS data;
August 18 is the latest complete snapshot used in the paper. The interval has
112 complete daily partitions after omitting the missing August 6 partition.
The scope, tuple key, service filter, deduplication, and interpretation
of operational exceptions remain fixed across the series.

In the 27-scenario scope, raw and actionable tuple break rates fall from
32.52\% to 19.82\% and from 20.12\% to 8.55\%, respectively. Raw and actionable
QPS-weighted rates fall from 19.68\% to 9.76\% and from 6.90\% to 1.87\%.
Observed services increase from 2,276 to 2,587 ($+13.7\%$). Table
\ref{tab:appendix-production-burden} records the corresponding absolute
counts. Counting only calls to Go services, the raw and actionable tuple rates
fall from 26.98\% to 16.45\% and from 13.59\% to 5.93\%, and their
QPS-weighted counterparts fall from 4.08\% to 2.10\% and from 3.37\% to
1.46\%.

The content-feed graph moves in the same direction. Observed services increase
from 1,440 to 1,528 ($+6.1\%$). Its raw and actionable tuple rates fall from
29.76\% to 19.51\% and from 16.38\% to 7.48\%, and the QPS-weighted rates fall
from 20.53\% to 9.99\% and from 5.61\% to 1.61\%. Counting only calls to Go
services, the tuple rates fall from 25.86\% to 17.04\% and from 11.73\% to
6.03\%, and the QPS-weighted rates from 2.34\% to 1.44\% and from 2.18\% to
1.28\%. Raw broken QPS on this graph falls from 205.39 million of
1.000 billion to 113.19 million of 1.133 billion.

Averaged over the last 28 days of the window, every metric remains below its
mean over the first 28 days.
The 27-scenario actionable tuple rate averages 13.16\% against 20.45\%, a drop
of 7.29 percentage points, and each of the last 28 days is below the early
mean. The actionable QPS-weighted rate averages 3.16\% against 6.94\%.
The content-feed raw QPS-weighted rate averages 10.66\% against 19.84\%.
Traffic mix, topology, unrelated deployments, manual repairs, and
ordinary product evolution also change during this period, so the series is
used as production context rather than as a treatment estimate.

\begin{table}[t]
  \centering
  \setlength{\tabcolsep}{4pt}
  \begin{tabular}{@{}lrr@{}}
    \toprule
    & \textbf{Apr 28} & \textbf{Aug 18} \\
    \midrule
    Broken / observed tuples & 4,947 / 24,590 & 2,235 / 26,136 \\
    Broken QPS (millions) & 86.51 & 26.59 \\
    Total QPS (billions) & 1.254 & 1.422 \\
    \bottomrule
  \end{tabular}
  \caption{Counts behind the actionable break rates in the 27-scenario scope.}
  \label{tab:appendix-production-burden}
\end{table}

Table~\ref{tab:appendix-language-break-rates} reports the actionable tuple
break rate for each primary implementation language with at least 100 tuples
at both endpoints. All four language groups improve.

\begin{table}[t]
  \centering
  \begin{tabular}{@{}lrr@{}}
    \toprule
    \textbf{Language} & \textbf{April 28} & \textbf{August 18} \\
    \midrule
    Go & 13.59\% & 5.93\% \\
    C++ & 29.95\% & 14.09\% \\
    Python & 15.95\% & 8.04\% \\
    Java & 63.48\% & 58.33\% \\
    \bottomrule
  \end{tabular}
  \caption{Actionable tuple break rate by primary implementation language.}
  \label{tab:appendix-language-break-rates}
\end{table}

The rollout-aligned analysis starts from 424 call tuples with an explicit
post-deployment record that the propagation break is no longer observed. It
retains 240 tuples whose governance record
links to a merged application-repository repair and whose caller service has a
recorded production-deployment time after the merge; these tuples belong to 16
services. The analysis drops the other 184 tuples: 19 link only to a merged
shared-repository repair, 56 have a merge request that was never merged, and
109 carry the production record without any link to a merge request.

For each service, day 0 is its first successful production deployment recorded
after the linked repair was merged. The event confirms that a deployment
occurred after the merge,
but not that the exact repaired commit was included. The merge to rollout proxy delay
ranges from 0.12 to 14.20 days, with median 6.62 days and 90th percentile
11.22 days. The 240 measurement units are call tuples. Tuples within one service
can share code, root cause, merge request, and deployment, so uncertainty is
clustered by service.

Presence is calculated over daily observations, not as a single count divided
by 240. For a window of $d$ days, the denominator is the $240d$ expected tuple
observations, and the numerator is the number present in telemetry. The three
windows contain 3,312/3,360, 1,655/1,680, and 5,611/5,760 observations, yielding
98.57\%, 98.51\%, and 97.41\%, respectively.
Table~\ref{tab:appendix-event-windows} reports presence, tuple break rate, and
QPS-weighted break rate in each window. Table~\ref{tab:appendix-event-days}
lists selected event days.

\begin{table}[t]
  \centering
  \begin{tabular}{@{}lrrr@{}}
    \toprule
    \textbf{Window} & \textbf{Presence} & \textbf{Tuple} & \textbf{QPS} \\
    \midrule
    Days $-14$ to $-1$ & 98.57\% & 90.46\% & 18.13\% \\
    Days 0--6 & 98.51\% & 8.70\% & 0.60\% \\
    Days 7--30 & 97.41\% & 4.69\% & 0.42\% \\
    \bottomrule
  \end{tabular}
  \caption{Pooled presence and break rates for the 240 target tuples.}
  \label{tab:appendix-event-windows}
\end{table}

\begin{table}[t]
  \centering
  \begin{tabular}{@{}lrrr@{}}
    \toprule
    \textbf{Day} & \textbf{Presence} & \textbf{Tuple} & \textbf{QPS} \\
    \midrule
    $-14$ & 100.00\% & 97.08\% & 20.25\% \\
    $-7$ & 100.00\% & 89.58\% & 18.01\% \\
    $-1$ & 99.58\% & 79.92\% & 13.66\% \\
    $0$ & 98.75\% & 16.46\% & 0.78\% \\
    $1$ & 98.75\% & 8.02\% & 0.57\% \\
    $6$ & 98.33\% & 7.63\% & 0.57\% \\
    $7$ & 98.33\% & 7.63\% & 0.55\% \\
    $14$ & 97.50\% & 3.85\% & 0.49\% \\
    $21$ & 97.50\% & 4.27\% & 0.34\% \\
    $30$ & 92.08\% & 4.52\% & 0.42\% \\
    \bottomrule
  \end{tabular}
  \caption{Selected event-day presence and break rates.}
  \label{tab:appendix-event-days}
\end{table}

Each of the 16 services has a lower break rate after deployment than in its
pre-deployment window. Table~\ref{tab:appendix-event-study} reports unweighted
service means and bootstrap intervals clustered by service. A
\emph{percentage-point change} is the direct difference between two percentage
rates. For example, a rate that falls from 90\% to 10\% decreases by 80
percentage points; this differs from its relative decrease of 88.9\%.
Even the lower endpoint of each 95\% interval shows a break-rate reduction of
more than 60 percentage points. In contrast, both 95\% intervals for presence
include no change. Target calls disappearing from telemetry therefore do not
explain the reduction.

\begin{table}[t]
  \centering
  \begin{tabular}{@{}lrr@{}}
    \toprule
    \textbf{Metric} & \textbf{Days 0--6} & \textbf{Days 7--30} \\
    \midrule
    Break-rate reduction & $75.12$ & $80.50$ \\
    95\% interval & $[61.69,87.16]$ & $[70.04,89.94]$ \\
    Presence & $+0.22$ & $-1.42$ \\
    95\% interval & $[-2.28,+2.77]$ & $[-6.20,+2.38]$ \\
    \bottomrule
  \end{tabular}
  \caption{Service-level break-rate reductions and presence changes relative
  to the pre-deployment window, in percentage points.}
  \label{tab:appendix-event-study}
\end{table}

\subsection{Workflow Adoption and Remaining Cases}
\label{sec:appendix-adoption}

Each mode keeps its own governance records, and the two label their cases
differently, so we report them separately rather than pooling them. The
interactive records contain 1,667 tracked broken calls and 144
merge requests across application and shared repositories. The batch records
contain 3,057 cases: 1,460 inside the scope operators gave the
campaign, 1,276 outside it, and 321 unlabeled. Both contain only the
items present in the governance tables at the snapshot, not every production
call or every raw detector output. One merge request can link to several call
records.

Of the interactive records, 852 are completed, 169 deferred, 147 have a draft
merge request, 114 are queued, 65 inconclusive, 37 analysis-complete, 24
pending, nine blocked, two abandoned, and 248 blank. In total, 830 of 1,667 ($49.8\%$) are marked
fixed. Explicit
production verification exists for 442 records: 424 positive and 18 negative
($424/442=95.9\%$ among labeled outcomes). The other
1,225 records have no such label. Among the batch records, 280 deployed cases
carry a post-deployment result: 192 repaired and 88 not
($192/280=68.6\%$).

All 144 merge requests come from interactive mode.
Table~\ref{tab:appendix-mr-funnel} follows them from proposal to deployment,
grouped by the repository each repair modifies. Deployment is recorded only
for application repositories, because a shared-library repair reaches
production through its consumers. The application requests that are not merged
are 31 draft, 10 open, and one approved; the shared requests that are not
merged are 13 draft, 11 open, seven rejected, three closed, and one left on an
old branch. The agent audited 112 of the 144 requests
($77.8\%$) and passed 98 of them ($87.5\%$). Audit and owner-status fields are
both populated for 83 requests: 60 pass both, 13 pass the audit while an owner
has yet to review them or lacks access, and six the audit called a wrong
fix but an owner passed. The remaining four have other combinations of audit
and review outcomes. Another 30 requests that owners passed or merged have
no audit.

\begin{table}[t]
  \centering
  \begin{tabular}{@{}lrrr@{}}
    \toprule
    & \textbf{Application} & \textbf{Shared} & \textbf{Total} \\
    \midrule
    Merge requests & 78 & 66 & 144 \\
    Owner reviewed & 63 & 51 & 114 \\
    Merged & 36 & 31 & 67 \\
    Marked online & 25 & & 25 \\
    Passed or merged & 48/63 & 48/51 & 96/114 \\
    Passed audit & 54/62 & 44/50 & 98/112 \\
    \bottomrule
  \end{tabular}
  \caption{Interactive-mode merge requests, by the repository the repair
  modifies.}
  \label{tab:appendix-mr-funnel}
\end{table}

Merging a shared-library repair does not repair its callers, because each
caller must still move to the new version, so \sys{} records closure at the
affected call. Among 123 calls linked to shared or dependency repairs, 112 have
an identified repair ($91.1\%$). A separate field tracks downstream adoption: 90
are complete, 23 have a draft merge request, nine are deferred, and one is
abandoned. The 123 calls span 33 caller groups and 47 shared repositories.

Table~\ref{tab:appendix-owner-feedback} reports owner responses. The source
contained 124 analyzable records; 16 were routing-only reassignment or owner
mismatch and are excluded, leaving 108 product-relevant records.

\begin{table}[t]
  \centering
  \begin{tabular}{@{}lrr@{}}
    \toprule
    \textbf{Owner response} & \textbf{Count} & \textbf{Share} \\
    \midrule
    Explicit positive or already handled & 30 & 27.8\% \\
    Default-positive operational closure & 46 & 42.6\% \\
    Process action or further validation & 16 & 14.8\% \\
    Question about mechanism or impact & 9 & 8.3\% \\
    Explicit negative or risk concern & 3 & 2.8\% \\
    Other or unclear & 4 & 3.7\% \\
    \textbf{Total} & \textbf{108} & \textbf{100.0\%} \\
    \bottomrule
  \end{tabular}
  \caption{Product-relevant owner responses to proposed repairs.}
  \label{tab:appendix-owner-feedback}
\end{table}

The 1,276 batch records outside the campaign scope carry a reason for
exclusion. The largest reasons are sidecar-originated calls (482), no static
execution path (222), no released fixed dependency (210), insufficient break
evidence (116), and a required breaking dependency upgrade (114). These five
account for 89.66\% of the cohort. The remaining 132 are
unresolved dependency attribution (62), other evidence gaps (26), out-of-scope
repair points (19), unresolved dynamic dispatch (18), and calls outside the
campaign target (7).

\subsection{Repeated Measurement of the Content-Feed Graph}
\label{sec:appendix-frontier}

The six discovery rounds are scoped to the content-feed graph
in one production region. At the production snapshot, this graph contains
16,251 calls, 1,528 visible services, and approximately 1.13 billion
successful QPS. It is part of the 27-scenario production scope,
not a platform-wide fleet.

Each round measures one hour of telemetry for this graph, selects services
with an observed propagation break and measured traffic above five QPS, excludes
services selected in earlier rounds, routes eligible cases
for repair, and repeats the measurement after production changes can take
effect. The rounds identify
586, 637, 414, 157, 51, and 37 newly visible candidate services. Their
cumulative counts are 586, 1,223, 1,637, 1,794, 1,845, and 1,882.

The procedure intentionally reports first-seen services, and the exported
lists contain no exact repeats across rounds. The reduction from the round-2
peak to round 6 is 94.19\%. This means that the operational discovery workload
falls from hundreds of services per round to 51 and then 37, making subsequent
repair waves manageable as routine follow-up. A separate stopping rule requires
new services to remain below 1\% of the cumulative set for two rounds. Rounds 5
and 6 contribute 2.76\% and 1.97\%, respectively.

\subsection{Durability Rule}
\label{sec:appendix-durability}

This analysis follows the same 240 target tuples across 16 services. Each tuple
uses its service's rollout-proxy date, and days 0--6 form a rollout grace
period. A recurrence requires telemetry to report the propagation break on two
consecutive observed days. We count a break as nonrecurring only if its call
tuple remains observable in the final seven days of the window; otherwise the
analysis omits it rather than interpreting absence as repair. We separately report recurrence among calls whose repair
should preserve request causality, excluding expected periodic,
self-initiated, and uncontrolled refresh work.

At 30 days, four of 240 calls are omitted because they are no longer
observable. Fifteen of the remaining 236 reappear ($6.36\%$), but none of the
82 observable request-continuation calls recur. At 60 days, 56 calls are
omitted because they are no longer observable; the same 15 calls remain the
only ones that reappear, now among 184 observable calls ($8.15\%$), and none
of the 78 request-continuation calls recur. Fourteen of the 15 are
periodic or self-initiated and one is an uncontrolled dependency refresh.

For the 82 request-continuation calls observed through 30 days and the 78
observed through 60 days, zero recurrences yield approximate 95\% upper bounds
of $3/82=3.66\%$ and $3/78=3.85\%$ on the
recurrence rate. This ``rule of three'' is the standard zero-event estimate:
when no event occurs in $n$ observations, $3/n$ is an approximate upper bound
on the underlying event rate at 95\% confidence.

\subsection{Operational Efficiency}
\label{sec:appendix-efficiency}

These measurements come from the agent's own run logs, read on August 15,
2026. They cover the 16 caller services for which the agent completed at least
one repair; status checks and sessions that produced no repair work are
excluded. Nine of the 16 repairs ended with a passing audit. The 16 services
took 43 agent sessions and 9.99 hours of agent time in total. A service took
21.95 minutes at the median, and half of the services fell between 12.87 and
30.89 minutes.

The logs also report how many tokens each session exchanged with the model, measured against a fixed baseline recorded before the run. For the
services whose sessions are all accounted for, repair used 169.75 million
tokens: 169.20 million sent to the model and 0.55 million returned. Most of
what the agent sent was context the model had already seen and served from
its cache, 153.60 million tokens or 90.78\% of the input. A service used
10.04 million tokens at the median, and half fell between 7.86 and 16.06
million.

\subsection{Rollback Case Study}
\label{sec:appendix-rollback}

During a rollout on June 11, 2026, a service owner saw the reported request
latency rise by about $300\%$ and rolled it back. The rollout had also
upgraded a tracing library that the service pulled in through another
dependency. That library measured its own periodic calls as if they
were incoming requests, so their durations landed in the same latency metric
as real traffic. A later change removed the library and recorded the periodic
calls under a separate span. Test environments that serve no traffic then
stopped reporting these measurements, which confirmed that the metric had
changed rather than the time spent serving requests.

\subsection{How the Audit Compares with Production Outcomes}
\label{sec:appendix-acceptance}

Two case sets ask whether the audit verdict agrees with what production later
showed. The first is a panel of 50 cases drawn from 20 caller services,
deliberately sampled to include different outcomes: 20 repaired calls that
continue a request, 10 repaired periodic or self-initiated calls, 10 calls
that were deployed but not repaired, and 10 that were deferred or left
inconclusive. Production later recorded an outcome for 40 of them, 30 positive
and 10 negative. Because we chose that mix, the ratio describes the panel and
not the fleet. The agent audited 39 of the 50 repairs, and an engineer
reviewed all 50.

The second set records each audit verdict before the production outcome is
known. Across 27 repairs, 21 passed the audit, two failed it, and four lacked
enough evidence for a verdict. All 27 later recovered in
production. The audit therefore agreed with production on 21 of the 23
repairs it ruled on, and its two rejections were conservative. No repair in
this set failed, so the set leaves open how often an audit-passing repair
fails in production.

\end{document}